\documentclass[twocolumn,showpacs,prX,nofootinbib,nobibnotes]{revtex4}

\usepackage{graphicx}
\usepackage{dcolumn}
\usepackage{bm}

\usepackage{anysize}
\usepackage{graphicx}
\marginsize{1.8 cm}{1.5 cm}{0.5 cm}{3.0 cm}

\usepackage{graphicx}
\usepackage{dcolumn}
\usepackage{bm}
\usepackage{amsmath}
\usepackage{amssymb}
\usepackage{mathrsfs}
\usepackage[latin1]{inputenc}
\usepackage{amsmath}

\newcommand{\dd}{{\rm d}}

\newcommand{\sd}{Schr\"{o}dinger }

\newcommand{\C}{\mathbb{C}}

\newcommand{\U}{\mathcal{U}}

\newcommand{\hil}{\mathcal{H}}

\newcommand{\tr}{{\rm Tr}}

\begin {document}


\title{Orthogonal measurement-assisted quantum control}

\author{Raj Chakrabarti}
\affiliation{Department of Chemistry, Princeton University,
Princeton, New Jersey 08544, USA}
\email{rajchak@Princeton.Edu}

\author{Rebing Wu}
\affiliation{Department of Chemistry, Princeton University,
Princeton, New Jersey 08544, USA}

\author{Herschel Rabitz}
\affiliation{Department of Chemistry, Princeton University,
Princeton, New Jersey 08544, USA}

\date{24 August 2007}

\begin{abstract}

Existing algorithms for the optimal control of quantum observables
are based on locally optimal steps in the space of control fields,
or as in the case of genetic algorithms, operate on the basis of
heuristics that do not explicitly take into account details
pertaining to the geometry of the search space. We present globally
efficient algorithms for quantum observable control that follow
direct or close-to-direct paths in the domain of unitary dynamical
propagators, based on partial reconstruction of these propagators at
successive points along the search trajectory through orthogonal
observable measurements. These algorithms can be implemented
experimentally and offer an alternative to the adaptive learning
control approach to optimal control experiments (OCE). Their
performance is compared to that of local gradient-based control
optimization.

\end{abstract}


\maketitle

\section{Introduction}\label{intro}

The optimal control of quantum dynamics is receiving increasing
interest due to widespread success in laboratory and computational
experiments across a broad scope of systems. With these promising
results it becomes imperative to understand the reasons for success
and to develop more efficient algorithms that can increase objective
yields to higher quality. In the computational setting, the expense
of iteratively solving the Schrodinger equation necessitates faster
algorithms for the search over control field space if these methods
are to be routinely employed for large-scale applications. In the
laboratory setting, although closed-loop methodologies have
encountered remarkable success, the search algorithms currently used
do not typically attain yields as high as those that can be achieved
using computational algorithms.

Recently, significant strides have been made towards establishing a
foundation for the systematic development of efficient OCT
algorithms based on the observation that the landscape traversed by
search algorithms in the optimization of quantum controls is not
arbitrarily complicated, but rather possesses an analytical
structure originating in the geometry of quantum mechanics
\cite{Raj2007}. This structure should allow not only rationalization
of the comparative successes of previous quantum control
experiments, but also analytical assessment of the comparative
efficiencies of new algorithms.

Prior work established important features of these landscapes, in
particular, their critical topologies \cite{RabMik2005,Mike2006a}.
The critical points of a control landscape correspond to locally
optimal solutions to the control problem. The most common objective
in quantum optimal control is maximization of the expectation value
of an observable. The landscape corresponding to this problem was
shown to be almost entirely devoid of local traps, i.e., the vast
majority of local suboptima are saddles, facilitating the
convergence of local search algorithms. The number of local
suboptima, as well as the volumes of these critical regions were
calculated. However, the relationship between the topology of
quantum control landscapes and their geometry, which would dictate
the behavior of global search algorithms, was not explored.

Thus far, optimal control algorithms for quantum observables have
not exploited the geometry of quantum control landscapes, which may
simplify control optimization compared to that for classical
systems. Indeed, the majority of quantum optimal control algorithms
to date have aimed at optimizing an objective functional, such as
the expectation value of an observable operator, directly on the
domain of time-dependent control fields $\varepsilon(t)$. Typical approaches
to quantum control optimization use the information in the
measurement of a single quantum observable to guide the search for
optimal controls; the simplest approach is to randomly sample single
observable expectation values at various points over the landscape
and use genetic algorithms (GA) to update the control field. A recent
experimental study \cite{Roslund2007} demonstrated at least a
two-fold improvement in optimization efficiency through the use of
local gradient algorithms rather than GA, but the important question
remains as to whether global algorithms for quantum control that are
not "blind" like GA can be implemented in an experimental setting.

When control optimization seeks to optimize the expectation value of
an observable by following, e.g., local gradient information on the
domain of controls, the geometry of the underlying space of quantum
dynamical propagators $\U(N)$ is not explicitly exploited. In
particular, optimal control algorithms that are based on locally
minimizing an objective function on the domain of control fields do
not follow globally optimal paths in $\U(N)$. This approach tends to
convolute the properties of the map between control fields and
associated unitary propagators with the properties of the map
between unitary propagators and associated values of the objective
function.

An alternative approach to observable maximization is to first solve
numerically for the set of unitary matrices $U$ that maximize the
expectation value $\tr(U\rho(0)U^{\dag}\Theta)$ of the observable
$\Theta$, and then to determine a control field $\varepsilon(t)$ that produces
that $U$ at time $t=T$ \cite{Khaneja2001,Khaneja2002a}.  For
restricted Hamiltonians in low dimensions, analytical solutions for
the optimal control field $\varepsilon(t)$ have been shown to exist. However,
to date, numerical algorithms for the optimization of unitary
propagators in higher dimensions have operated solely on the basis
of local gradient information, such that the global geometry of
$\U(N)$ is again not exploited.

The variational problems of optimal control theory admit two types
of minimizers. Denoting the cost functional by $J$, according to the
chain rule,
$$\frac{\delta J}{\delta \varepsilon(t)} = \frac{\dd J}{\dd U} \cdot
\frac{\delta U}{\delta \varepsilon(t)}.$$ The first type of
minimizer corresponds to those control Hamiltonians that are
critical points of the control objective functional, but are not
critical points of the map between control fields and associated
dynamical propagators (i.e., points at which $\frac{\dd J}{\dd
U}=0$, while the Frechet derivative mapping from the control
variation $\delta \varepsilon(t)$ to $\delta U(T)$ at $t=T$ is
surjective).
The second type corresponds to critical points of the latter map
(i.e., points at which the mapping from $\delta \varepsilon(t)$ to
$\delta U(T)$ is not locally surjective)
\cite{Wu2007} \footnote{The local surjectivity of $\varepsilon(t)
\rightarrow U(T)$ has important connections to the controllability
of the quantum system \cite{Raj2007}}. Critical points of the first
type, which are referred to as kinematic critical points or normal
extremal controls, are either global optima or saddle points, but
never local traps\cite{RabMik2005,Mike2006a}. Recent work in quantum
optimal control theory suggests that the critical points of the map
$\varepsilon(t) \rightarrow U(T)$, called abnormal extremal
controls, are particularly rare (i.e., there are generally fewer
critical points compared to classical control problems)
\cite{Wu2007,Raj2007}.

Because the objective function $J$ is a complete function of $U(T)$,
the maximal achievable optimization efficiency is ultimately
determined by the properties of the map between control fields and
unitary propagators, $\varepsilon(t) \rightarrow U(T)$. Irrespective
of the corresponding observable expectation value, updating the
control field to produce a unitary propagator that is close to the
current propagator will typically be computationally inexpensive.
Therefore, following a direct route in the space of unitary
propagators is expected to be more efficient in quantum control than
following a gradient flow on the space of objective function values
that maps to a longer path in $\U(N)$. As will be shown, the
scarcity of critical points of the map $\varepsilon(t) \rightarrow
U(T)$ in quantum control problems implies that it is surprisingly
simple to track arbitrary paths in $\U(N)$ during optimization, at
least for certain families of Hamiltonians. However, it is not
uncommon to encounter regions of $\U(N)$ where numerically, the
relevant differential equations are ill-conditioned. The ability to
selectively avoid such singular regions, which correspond to
abnormal extremals, is desirable. One way to achieve this goal is to
constrain the search trajectory to only roughly follow a
predetermined path in $\U(N)$.

In this paper, we develop globally efficient algorithms for the
optimization of quantum observables that exploit the geometry of
$\U(N)$ by \textit{approximately} following a predetermined path in
the space of quantum dynamical propagators. This approach to
globally efficient quantum control optimization is based on making a
partial tomographic set of measurements at various steps along the
search trajectory. A complete tomographic set of observations is a
set that is adequate for the estimation of all the $N^2$ parameters
of the unitary propagator $U(T)$ \cite{Hradil2003,Lidar2007}; a partial tomographic set
reconstructs only a subset of these parameters. The goal of this
approach is to reap the benefits of unitary matrix tracking without
encountering the associated singularities. As such, the approach
attempts to leverage the methodologies of quantum statistical
inference \cite{Malley1993} in order to reduce the search effort
involved in solving quantum control problems.

These experimentally-implementable algorithms for
quantum control optimization can be simulated by employing a generalization of the
diffeomorphic homotopy tracking methodology D-MORPH
(diffeomorphic modulation under observable response-preserving
homotopy) \cite{Rothman2005,Rothman2006a,Rothman2006b}. In contrast
to observable-preserving diffeomorphic tracking, the orthogonal
observable tracking algorithm developed and applied here identifies
parametrized paths $\varepsilon(s,t)$ that follow a given predetermined
trajectory through $\U(N)$. In both cases, a denumerably infinite
number of solutions exist to the tracking differential equations;
paths $\varepsilon(s,t)$ that optimize desirable physical features of the
control field can be tuned through the choice of an auxiliary free
function. We will show that a primary difference between scalar and
vector observable tracking is that the trajectory followed in
$\U(N)$ in the former case is highly sensitive to changes in the
system Hamiltonian, whereas the $U$-trajectory followed in the
latter case can be rendered largely system independent by employing
a larger set of orthogonal observables. This suggests that, besides
its usefulness as an optimization algorithm, orthogonal observable
tracking can reveal universal features underlying the computational
effort involved in quantum optimal control searches across diverse
systems.

In addition, we compare the trajectories in $\U(N)$ followed by
standard OCT gradient-following algorithms with those that track
optimal paths in the dynamical group, for various Hamiltonians, in
order to determine how the geometry of the underlying space affects
the convergence of experimental and computational control
optimizations that exploit only local gradient information. In so
doing, we will show that there exists a special relationship between
the gradient flow on $\varepsilon(t)$ and a particular (global) path $U(s)$ in
the domain of unitary propagators, namely the gradient flow of the
objective on $\U(N)$, which offers insight into the convergence
properties of the former.

\section{Quantum optimal control gradient flows}\label{gradientflows}

Local algorithms for quantum optimal control, whether numerical
(OCT) or experimental (OCE), are typically based on the gradient of
the objective function. In this section, we review the properties of
the gradient for quantum observable expectation value maximization, and dissect
these properties into system(Hamiltonian)-dependent and universal
system-independent parts.

A generic quantum optimal control cost functional can be written:

\begin{multline}\label{OCT}
J =\Phi(U(T), T)-\\
\textmd{Re}\left[\tr\int_{0}^T\left\{\left(\frac{\partial U(t)}{\partial t} +
\frac{i}{\hbar}H(\varepsilon(t))U(t)\right)\beta(t)\right\}dt\right]-\\
\lambda\int_{0}^T \mid \varepsilon(t)\mid^2 \dd t
\end{multline}
where $H$ is the total Hamiltonian, $\beta(t)$ is a Lagrange
multiplier operator constraining the quantum system dynamics to obey
the \sd equation, $\varepsilon (t)$ is the time-dependent control
field, and $\lambda$ weights the importance of the penalty on the
total field fluence. Solutions to the optimal control problem
correspond to $\frac{\delta J}{\delta \varepsilon(t)} = 0$.  The
functional $\Phi$, which we refer to as the objective function, can
take various forms. The most common form of $\Phi$ is the
expectation value of an observable of the system:
$$\Phi(U) = \tr(U(T){\rho(0)}U^{\dag}(T)\Theta)$$
where $\rho(0)$ is the initial density matrix of the system and
$\Theta$ is an arbitrary Hermitian observable operator
\cite{Mike2006a}.

An infinitesimal functional change in the Hamiltonian $\delta H(t)$
produces an infinitesimal change in the dynamical propagator
$U(t,0)$ as follows:
$$\delta U(t,0) = - \frac {i}{\hbar} \int_0^t U(t,t') \delta H(t')
U(t',0)dt'$$ where $\delta H(t) = \triangledown_\varepsilon H(t)
\cdot \delta \varepsilon(t)$. The corresponding change in $\Phi$ is then given
by
$$\delta \Phi = -\frac {i}{\hbar} \int_0^T
\tr(~[\Theta(T),U^{\dag}(t,0)\delta H(t)U(t,0)~]\rho(0))dt,$$ where
$\Theta(T) \equiv U^{\dag}(T){\rho(0)}U(T)$. In the special case of
the electric dipole approximation, the Hamiltonian assumes the form
$$H(t) = H_0 - \mu \cdot \varepsilon(t)$$
where $H_0$ is the internal Hamiltonian of the system and $\mu$ is
its electric dipole operator. $\mu(t)$ is given by
$$\mu(t) \equiv U^{\dag}(t,0)\mu U(t,0) = i \hbar M(t)$$
where $M(t) \equiv \frac {i} {\hbar} U^{\dag}(t,0) \triangledown_{\varepsilon}
H(t) U(t,0)$. Within the electric dipole approximation, the gradient
of $\Phi$ is \cite{HoRab2007a}:
\begin{multline}\label{grad1}
\frac{\delta \Phi}{\delta \varepsilon(t)} =
-\frac{i}{\hbar}\tr\{\left[\Theta(T),\mu(t)\right]\rho(0)\}\\
= \frac{i}{\hbar}\sum_i
\rho(0)\langle i|\Theta(T)\mu(t)-\mu(t)\Theta(T)|i\rangle \\
= \frac {i}{\hbar} \sum_{i=1}^n p_i \sum_{j=1}^N \Big[\langle
i|\Theta(T)|j\rangle \langle j|\mu(t)|i\rangle -\langle
i|\mu(t)|j\rangle \langle j|\Theta(T)|i\rangle \Big]
\end{multline} where the initial density matrix is given as $\rho(0)=\sum_{i=1}^n
p_i |i\rangle\langle i|, p_1 > ...> p_n > 0, \quad \sum_{i=1}^n p_i
= 1.$ The assumption of local surjectivity of $\varepsilon(t)
\rightarrow U(T)$ implies that the functions $\langle
i|\mu(t)|j\rangle $ are $N^2$ linearly independent functions of
time. The functions
\begin{multline}\label{gradbasis}
\langle i|U^{\dag}(T) \Theta U(T)|j\rangle \langle j|U^{\dag}(t)\mu U(t)|i\rangle ~ -\\
\langle i|U^{\dag}(t)\mu U(t)|j\rangle \langle j|U^{\dag}(T)\Theta U^{\dag}(T)|i\rangle
\end{multline}
therefore constitute natural basis functions for the gradient on
the domain $\varepsilon(t)$. We are interested in the global behavior of the
flow trajectories followed by these gradients, which are the
solutions to the differential equations
\begin{equation}\label{Egrad}%
\frac{d\varepsilon(s,t)}{ds}= \triangledown \Phi_\varepsilon(\varepsilon(t))
=\alpha \frac {\delta \Phi(s,T) }{\delta \varepsilon(s,t)}
\end{equation}
where $s > 0$ is a continuous variable parametrizing the algorithmic
time evolution of the search trajectory, and $\alpha$ is an
arbitrary positive constant that we will set to 1. The existence of
the natural basis (\ref{gradbasis}) indicates that these flow
trajectories evolve on a low-dimensional subspace of $\varepsilon(t)$.
However, the gradient flow equations cannot be integrated
analytically for arbitrary internal Hamiltonians $H_0$, precluding a
deeper understanding of the global dynamics of the search process.
In fact, these dynamics do not have universal
(Hamiltonian-independent) properties. The explicit path followed by
the search algorithm on $\varepsilon(t)$ depends on the solution to
the Schrodinger equation for the particular system Hamiltonian and
cannot be expressed analytically.

Because the objective functional $\Phi$ is explicitly a function of
$U(T)$, any universal properties of the global geometry of the
search dynamics must be investigated on this domain. These search
dynamics are governed by the gradient flow of $\Phi$ on the domain
$\U(N)$, given by
$$\frac{dU}{ds} = \triangledown \Phi(U).$$
The tangent space of $\U(N)$ at any element $U \in
\U(N)$ is
$$T_U \U(N)=\{U \Omega | \Omega^{\dag} = -\Omega, \quad \Omega
\in \C^{N\times N}\},$$ where $\Omega$ is an arbitrary skew-Hermitian matrix,
and the directional derivative for a
function $\Phi$ defined on $\U(N)$ is
\begin{eqnarray*}
D \Phi_U(U \Omega) \equiv  \tr
\left((\triangledown\Phi(U))^{\dag}U\Omega\right).
\end{eqnarray*}
The directional derivative of the objective functional $\Phi$ along
an arbitrary direction $U \Omega$ in $T_U \U(N)$ can then be written
\begin{eqnarray*}
D\Phi_{U}(U\Omega) &=& \tr\left(U^{\dag}\Theta U \Omega \rho(0) + (U
\Omega)^{\dag}\Theta U \rho(0)\right) \\
&=& \tr\left([\rho(0),U^{\dag}\Theta U]
\Omega\right)
\end{eqnarray*}
allowing us to identify the gradient of $\Phi$ on $\U(N)$ as
$$\triangledown \Phi = -U[\rho(0),U^{\dag}\Theta U]=[\Theta,U\rho(0) U^{\dag}]U.$$
Therefore, the equations of motion for the gradient flow lines of
objective functional $\Phi$ are
\begin{equation}\label{Uflow}
\frac{dU}{ds} = [\Theta,U\rho(0)U^{\dag}]U = -U \rho(0) U^{\dag}\Theta U + \Theta U\rho(0).
\end{equation}
In section \ref{integration} below, we integrate these equations to obtain the
trajectories $U(s)$ followed by gradient algorithms on $\U(N)$ over algorithmic time
$0\leq s< \infty$.

The essential question arises as to the relationship between the
gradient flow on $\varepsilon(t)$ and that on $\U(N)$. The gradient on $\varepsilon(t)$ is related to the gradient on
$\U(N)$ through
\begin{equation}\label{chain}%
\frac{\delta \Phi}{\delta \varepsilon(t)}=\sum_{i,j}\frac{\delta
U_{ij}}{\delta \varepsilon(t)}\frac{\dd \Phi}{\dd U_{ij}}.
\end{equation}
Now suppose that we have the gradient flow of $\varepsilon(s,t)$ that
follows (\ref{Egrad}) and let $U(s)$ be the projected trajectory on the
unitary group $\U(N)$ of system propagators at
time $T$, driven by $\varepsilon(s,t)$. The algorithmic time derivative of $U(s)$
is then
\begin{equation}\label{Us}
  \frac{\dd U_{ij}(s)}{\dd s}=  \int_0^T \frac{\delta U_{ij}(s)}{\delta \varepsilon(s,t)}\frac{\partial \varepsilon(s,t)}{\partial
  s} \dd t
\end{equation}
which, combined with (\ref{Egrad}) and (\ref{chain}), gives
\begin{equation}\label{dot Us}
     \frac{\dd U_{ij}(s)}{\dd s}=\int_0^T \frac{\delta U_{ij}(s)}{\delta \varepsilon(s,t)}\sum_{p,q}\frac{\delta U_{pq}(s)}{\delta \varepsilon(s,t)}\frac{\dd \Phi}{\dd U_{pq}} \dd t.
\end{equation}
It is convenient to write this equation in vector form, replacing
the $N \times N$ matrix $U(s)$ with the $N^2$ dimensional vector
$\textbf{u}(s)$:
\begin{multline}\label{Gmat}
\frac{\dd \textbf{u}(s)}{\dd s} =\left[\int_0^T
\frac{\delta\textbf{u}(s)}{\delta
\varepsilon(s,t)}\frac{\delta{\textbf{u}^T(s)}}{\delta \varepsilon(s,t)}\dd
t\right]\triangledown \Phi[\textbf{u}(s)] \mathrel{\mathop:}= \\
\textmd{G}[\varepsilon(s,t)]\triangledown \Phi[\textbf{u}(s)]
\end{multline}
where the superscript $T$ denotes the
transpose. Thus the projected trajectory from the space of control
field is different from that driven by the gradient flow in the
unitary group:
\begin{equation}\label{gradient_U}
\frac{\dd U(s)}{\dd s}=\nabla \Phi[U(s)].
\end{equation}
This relation implies that the variation of the propagator in
$\U(N)$ caused by tracking the gradient flow in the space of control
field is Hamiltonian-dependent, where the influence of the
Hamiltonian is all contained in the $N^2$-dimensional symmetric
matrix $\textmd{G}[\varepsilon(s,t)]$.

\section{Unitary matrix flow tracking}\label{unitarytracking}

The matrix $\textmd{G}\left[\varepsilon(s,t)\right]$ in equation (\ref{Gmat}) above indicates
that the convergence time for local gradient-based OCT algorithms
may vary greatly as a function of the Hamiltonian of the system.
Given the decomposition of the gradient into Hamiltonian-dependent
and Hamiltonian-independent parts, the natural question arises as to
whether the Hamiltonian-dependent part can be suppressed to produce
an algorithm whose convergence time will be (approximately) dictated
by that of the unitary gradient flow, irrespective of the system
Hamiltonian.

In order for the projected flow from $\varepsilon(t)$ onto $U(T)$ to
match the integrated gradient flow on $U(T)$, the quantity
$\frac{\partial{\varepsilon(s,t)}}{\partial s}$ that corresponds to
movement in each step must satisfy a generalized differential
equation:
\begin{equation}\label{gendiff}
\frac{\dd U(s)}{ds} = \int_0^T \frac{\delta U(s)}{\delta
\varepsilon(s,t)}\frac{\partial{\varepsilon(s,t)}}{\partial s}\dd
t=\triangledown \Phi\left[U(s)\right].
\end{equation}

In the dipole approximation, this relation becomes the following
matrix integral equation:
$$\int_0^T \mu (s,t)\frac{\partial
{\varepsilon(s,t)}}{\partial s}\dd t = U^{\dag}(s)\triangledown \Phi
\left[U(s)\right],$$ where $\mu(s,t) \equiv U^{\dag}(s,t)\mu
U(s,t).$ When $\Phi$ is the observable expectation value objective
function, we have
$$\int_0^T \mu(s,t) \frac{\partial \varepsilon(s,t)}{\partial s}\dd t =
-\left[\rho,U^{\dag}(s)\Theta U(s)\right].$$  On the basis of
eigenstates, the matrix integral equation is written
\begin{equation}\label{matint}
\int_0^T \mu_{ij}(s,t)\frac {\partial \varepsilon(s,t)} {\partial s}
\dd t = i\hbar\langle i|U^{\dag}(s,T)\triangledown \Phi
\left[U(s,T)\right]|j\rangle.
\end{equation} To solve this equation, we first
note that the flexibility in the choice of the representation of the
variation in $\varepsilon(s,t)$ allows us to expand it on the basis
of functions $\mu_{ij}(s,t)$, as
$$\frac {\partial \varepsilon(s,t)}{\partial s} = \sum_{i,j} x_{ij}
\mu_{ij}(s,t).$$ Inserting this expansion into the above equation
produces
\begin{multline}
\sum_{p,q} x_{pq}(s) \int_0^T \mu_{ij}(s,t) \mu_{pq}(s,t)
\dd t = \\
i\hbar \langle i|U^{\dag}(s,T)\triangledown \Phi\left[U(s,T)\right]|j\rangle.
\end{multline}
If we denote the correlation matrix $\textmd{G}(s)$ as
\begin{multline}
\textmd{G}_{ij,pq}(s) = \int_0^T \mu_{ij}(s,t)\mu_{pq}(s,t) \dd t = \\
\int_0^T \langle i |\mu(s,t)|j\rangle \langle p|\mu(s,t)|q\rangle \dd t,
\end{multline}
(as in eqn (\ref{Gmat}) above, but now specifically in the case of
the dipole approximation) and define
$$\Delta_{ij}(s) \equiv i\hbar \langle i|U^{\dag}(s,T)\triangledown
\Phi\left[U(s,T)\right]|j\rangle,$$ it can be shown
\cite{Dominy2007} that the matrix integral equation (\ref{matint})
can be converted into the following nonsingular $N^2$-dimensional
algebraic differential equation:
\begin{equation}\label{utrack}
\frac{\partial \varepsilon}{\partial s} =  f_s + \Big(v(\Delta) -
\alpha \Big)^T\textmd{G}^{-1}v({\mu(t)})
\end{equation}
where $f_s = f_s(t)$ is a "free" function resulting from the
solution of the homogeneous differential equation, the operator $v$
vectorizes its matrix argument and $\alpha\equiv\int_0^T v(\mu(t))
f_s \dd t$.

Solving this set of $N^4$ scalar differential equations requires that the $N^2 \times N^2$ matrix
$\textmd{G}$ is invertible. The invertibility of this matrix is equivalent to the claim that the map
$\varepsilon(t)\rightarrow U(N)$ between control fields and unitary propagators is
surjective, such that it is possible to reach any $U(s+1)$
infinitesimally close to $U(s)$ in a vanishingly small step.
Thus, a necessary condition for the existence of
a well-determined search direction is the full-rank of the
Jacobian, i.e., $$\textmd{rank} \frac {\delta U(s)}{\delta \varepsilon(s,t)}
= \textmd{dim}[\U(N)] = N^2$$ which is equivalent to the
requirement of local surjectivity of the map $\varepsilon(t) \rightarrow U(T)$.

The problem is undetermined because of the rank of
the matrix is lower than the number of variables to be solved,
which results from the fact that the optimal control problem
itself is undetermined with a multiplicity of solutions. Each
"free function" $f_s$ corresponds to a unique algorithmic step in
$\varepsilon(t)$; modulating this function allows for systematic exploration
of the set of functions $\varepsilon(s,t)$ that are compatible with the
gradient step on $U$ \cite{Dominy2007}.

As in the case of the gradient $\triangledown \Phi[\varepsilon(t)]$, the flow
on $\varepsilon(t)$ that tracks the $U$-gradient can be expressed in terms of
a set of maximally $N^2$ linearly independent functions of time, but
whereas the former is a unique functional derivative, the latter is
highly degenerate. The former are explicitly determined by the
functions $\mu(t)$, while the latter are underdetermined by these
functions; only $N^2$ linearly independent components of
$\frac{\partial \varepsilon(s,t)}{\partial s}$ are explicitly determined by
$\mu(t)$, the rest remaining unspecified.

Although the gradient step $\frac{\dd \varepsilon(s,t)}{\dd s}=\nabla
\Phi[\varepsilon(s,t)]$ is always locally the direction of fastest decrease in
the objective function at $\varepsilon(t)$, the path $\varepsilon(s,t)$ derived from
following this gradient has no universal (Hamiltonian-independent)
global geometry, since $\Phi$ is not explicitly a function of
$\varepsilon(t)$. It is known \cite{Mike2006a} that this path will not
encounter any traps during the search, but beyond this, the geometry
can be expected to be rugged and globally suboptimal. Unlike the
gradient $\triangledown \Phi[\varepsilon(t)]$, the algorithmic step above
follows the gradient flow on $\U(N)$ (in the limit of
infinitesimally small algorithmic time steps). The $N^2$ functions
$\mu(s,t)$ are calculated during the evaluation of $\triangledown
\Phi[\varepsilon(t)]$; hence, the computational overhead incurred by following
this flow corresponds to that needed to compute the $N^4$ elements
of $\textmd{G}(s)$ and invert the matrix, at each algorithmic time step.
This flow respects the geometric formulation of the optimal control
objective function in terms of $U(T)$ rather directly in terms of
$\varepsilon(t)$. As we shall show below, the global geometry of this path can
be completely determined analytically for objective function $\Phi$.
The functions $\mu(s,t)$ contain all relevant information about the
quantum dynamics, whereas the functions $\triangledown \Phi(U)$
contain complete information about the geometry of the search space.

Incidentally, matrix integral equation (\ref{matint}) can be reformulated to
provide further insight into the relationship between
$\varepsilon(t)$-gradient and $U(T)$-gradient flows. Equation (\ref{matint}) can be
rewritten
\begin{multline}\label{mu}
\int_0^T\langle \mu(s,t'),\mu(s,t)\rangle \frac{\dd \varepsilon(s,t')}{\dd s}
\dd t' =\\
\langle -i\hbar \left[\Theta(s,T),\rho\right],\mu(s,t)\rangle.
\end{multline} It can be shown
that if the time-dependent dipole operator $\mu(s,t)$ displays the
Dirac property
\begin{widetext}
$$\langle \mu(s,t'),\mu(s,t) \rangle = \sum_{i=1}^N\sum_{j\geq i}^N \left[\textmd{Re}(\langle i | \mu(s,t') |j \rangle)\textmd{Re}(\langle j | \mu(s,t') | i \rangle)+\textmd{Im}(\langle j | \mu(s,t') | i \rangle)\right]=\delta(t-t'),$$
\end{widetext}
the corresponding nonsingular initial value problem is
$$\frac{\dd \varepsilon(s,t)}{\dd s} \approx \langle -i\hbar \left[\Theta(s,T),\rho(0)\right],\mu(s,t)\rangle,$$
which is effectively identical to the $\varepsilon(t)$-gradient flow for
observable functional $\Phi$. As such, the extent to which condition
(\ref{mu}) is satisfied for a given Hamiltonian will determine the
faithfulness with which this flow tracks the $U(T)$-gradient.

Of course, a multitude of other flows could be substituted for the
RHS of equation (\ref{matint}). In section \ref{integration}, we
will integrate the $U(T)$-gradient flow and show that it does not
follow a globally optimal path. Since we are interested in global
optimality, we should choose a flow that follows the shortest
possible path from the initial condition to a unitary matrix that
maximizes the observable expectation value. It can be shown
\cite{Mike2006a} that a continuous manifold of unitary matrices $W$
maximizes $\Phi(T)$. These $W$s can be determined numerically by
standard optimization algorithms on the domain of unitary
propagators \cite{Brockett1991}. The shortest length path in $U(N)$
between $U(0)$ and an optimal $W$ is then the geodesic path that can
be parameterized as $U(s)=U(0)\exp(iAs)$ with $A =
-i\log(W^{\dag}U(0))$ where $\log$ denotes the complex matrix
logarithm with eigenvalues chosen to lie on the principal branch
$-\pi<\theta<\pi$. Thus, if we set $\Delta_{ij}(s) = \langle i|A|j
\rangle = \langle i|-i\log(W^{\dag}U(s)|j \rangle$, the tracking
algorithm will attempt to follow this geodesic path. Because this
choice of $A$ does not represent the gradient of an objective
function, the optimization will not converge exponentially to the
solution, but rather will continue past the target matrix $W$ unless
stopped \cite{Dominy2007}. On the other hand, it is in principle
possible to choose $A$ that results in the algorithm tracking the
same step in $\U(N)$, but at a rate that depends on algorithmic time
$s$. \footnote{In the case that the control system evolves on a
subgroup of $U(N)$, e.g. SU(N), the geodesic on that subgroup can be
tracked instead.}

Due to the nonlinearity of the differential equations above, errors
in tracking will inevitably occur, increasing the length of the
search trajectory beyond that of the minimal geodesic path (see
below). These errors will naturally be a function of the system
Hamiltonian. It is of interest to examine the dependence of
matrix flow tracking errors on the Hamiltonian by continuously
morphing the Hamiltonian during the optimization. This efficient
approach to Hamiltonian sampling will allow a more systematic
comparison of the efficiency of global OCT optimization with that of
local gradient-based OCT, which is expected to be much more
system-dependent.

Hamiltonian morphing can encompass changes in both the system's internal Hamiltonian
and the dipole operator. We assume these matrices can be written as functions
of the algorithmic step $s$ as $\hil(s) = \hil_0(s) + \mu(s)\varepsilon(s,t)$. Since we have

$$\frac{\partial \hil(s,t)}{\partial s} = \frac{\dd \hil_0(s)}{\dd
s} - \frac{\dd \mu(s)}{\dd s} \varepsilon(s,t) - \mu(s)\frac{\partial
\varepsilon(s,t)}{\partial s},$$ we can rewrite eqn (\ref{matint}) as
\begin{multline}\label{duds2}
\frac{\dd U(s)}{\dd s} = \int_0^T \dd t \big(a_0(s,t,T)\frac{\partial
\varepsilon(s,t)}{\partial s} ~+\\
a_1(s,t,T)\varepsilon(s,t)+a_2(s,t,T)\big) = 0
\end{multline}
where $a_0 = \mu(s,t)$, $a_1 =\frac {\dd \mu(s,t)}{\dd s}$ and $a_2
= \frac {\dd \hil_0(s)}{\dd s}$. Thus, if we define
$$b(s,T) \equiv - \int_0^T \big( \nu(a_1)\varepsilon(s,t) + \nu(a_2) \big) \dd t,$$ where $\nu$ denotes the $N^2$-dimensional
vectorized Hermitian matrix as above, we can rewrite the matrix integral equation (\ref{matint}) as
$$\int_0^T \mu(s,t) \frac{\partial \varepsilon(s,t)}{\partial s}\dd t =
-~[\rho,U^{\dag}(s)\Theta U(s)~]+b(s,T).$$
Therefore in the case of combined Hamiltonian morphing and unitary
tracking, the D-MORPH differential equation for the control field
becomes

\begin{equation}\label{UHamdiff}
\frac{\partial E}{\partial s} =  f_s + \Big(v(\Delta) + b(s,T) -
\alpha \Big)^T\textmd{G}^{-1}v({\mu(t)})
\end{equation}

Even if $\textmd{G}$ is invertible, it is possible that it is nearly
singular, resulting in large numerical errors during the solution to
the differential equation. It is convenient to assess the nearness
to singularity of $\textmd{G}$ by means of its condition number $C$, namely
the ratio of its largest singular value to its smallest singular
value, i.e., if $\textmd{G}^{-1}=V\left[\textmd{diag}(1/\omega_j)\right]U^T$,
$C=\frac{\max_j\omega_j}{\min_j\omega_j}$.

As mentioned, tracking errors can also originate due to the omission
of higher order functional derivatives, such as $\frac{\delta^2
U(T)}{\delta \varepsilon(t)^2}$, in equation (\ref{duds2}).  These may in
principle be large, even if the input-state map is surjective,
resulting in large numerical errors for finite step sizes. Since the
calculation of these higher derivatives is very expensive, we do not
employ them in the calculation of the algorithmic step. The
hypothesis that tracking globally optimal paths in $\U(N)$ is
typically more efficient than local optimization of the expectation
value of the target observable is equivalent to the assumption that
the second and higher order functional derivatives of the unitary
propagator with respect to the control field are relatively small,
but not negligible when attempting to traverse a large distance in
$\U(N)$ in a single control field iteration.

The computational expense of unitary matrix tracking increases
fairly steeply with system dimension. Since matrix inversion scales
as $N^2$, where $N$ is the dimension of the matrix, the cost of
inverting the $\textmd{G}$ matrix scales as $N^4$, where $N$ is the Hilbert
space dimension. By contrast, global observable expectation value
tracking, discussed in the next section, avoids this overhead, but
at the cost of being unable to specify precisely the unitary path
followed during optimization.

In order to test the hypothesis that the primary determinant of optimization efficiency
is the unitary path length to the target $W$, we compared the optimization
efficiencies of algorithms that follow a geodesic on the unitary
group versus a faster path on the domain of objective function
values that corresponds to a longer path in $\U(N)$. These results are presented
in section \ref{numerical}.

\section{Orthogonal observation-assisted quantum
control}\label{orthog}

Unitary matrix tracking has the distinct advantage that it can
directly follow an optimal path in the space of unitary propagators,
assuming the input-state map is surjective and the linear
formulation of the tracking equations above is a reasonable
approximation. However, it cannot be implemented experimentally
without expensive tomography measurements, and carries a
computational overhead that scales exponentially with system size.

Given the initial state $\rho(0)$ of the system, matrix elements of the
unitary operator $U(T)$ can be determined based on knowledge of the final
state $\rho(T) = U(T)\rho(0)U^{\dag}(T)$. $\rho(T)$ can be known
only if a so-called tomographically complete set of observables has
been measured sufficiently many times on identical copies of the
system to approximate the expectation value of each observable. Assuming $\rho(0)$
is nondegenerate, if such measurements are made at each step of the control optimization,
the unitary matrix tracking described above can be implemented (if $\rho(0)$ is
degenerate, the maximum number of $U(T)$ elements that can be reconstructed will be diminished,
as shown below). However, the cost of this procedure is very steep for large systems.
The natural question arises as to what sort of comparative benefit
in optimization efficiency can be accrued from measurement of a
limited number $m$ of (orthogonal) operators, where $m < N^2$.

Consider the case where $n \geq m$ distinct observables, denoted
$\Theta_1(T),...,\Theta_n(T)$ or $\{\Theta_k\}$, possibly linearly
dependent and not necessarily orthogonal, are measured at each step.
For simplicity, represent each of the Hermitian matrices as an
$N^2$-dimensional vector with real coefficients. Then by
Gram-Schmidt orthogonalization, it is always possible to construct
an orthogonal basis of $m$ linearly independent $N^2$-dimensional
vectors, $\Theta'_1(T),...,\Theta'_m(T)$ that (spans this set) - any
element of the set $\{\Theta_k\}$ can be expressed as a linear
combination of the basis operators in this set, i.e., for any $k$,
$\Theta_k = \sum_{i=1}^m c_{ik} \Theta'_i$. In other words, the
information obtained by measuring the expectation values of the set
$\{\Theta_k\}$ is equivalent to that obtained by measuring the
$\{\Theta'_i\}$, since for each $k$, $\langle \Theta_k \rangle =
\langle \sum_{i=1}^m c_{ik} \Theta'_i \rangle$. 
orthogonal bases of Hermitian operators are the Pauli (2-d) and
Gell-Mann (3-d) matrices.

As above, we restrict ourselves here to coherent quantum dynamics,
and additionally assume that a sufficient number of measurements of
each observable have been made to accurately estimate its
corresponding expectation value. Now consider the $m$ (scalar
functions of algorithmic time) $\{\langle \Theta_k(T,s) \rangle\}$
of expectation values for each observable corresponding to a desired
unitary track $U(T,s)$. Again, the information about the states of
the system $\rho(T,s)$ or equivalently, $U(T,s)$ contained in these
measurements is equivalent to that contained in the $m$ functions
$\{\langle \Theta'_i(T,s) \rangle\}$. Let us therefore represent
this information in the form of the $m$-dimensional vector
$\textbf{v}(T,s)$, where
$$\textbf{v}_i(T,s) \equiv \langle \sum_k c_{ki}\Theta'_i(T,s)
\rangle.$$ During control optimization, we are interested in
tracking these paths $\textbf{v}(T,s)$ in the vector space
$\textmd{V}$ that are consistent with the desired path $Q(T,s)$ in
$U(N)$.

The generalized differential equation (analogous to eqn
(\ref{gendiff}) ) that must be satisfied in order to simultaneously
track these paths is:
\begin{multline}
\frac{\dd \textbf{v}(T,s)}{\dd s}=\int_0^T \frac{\delta \textbf{v}(T,s)}{\delta \varepsilon(s,t)}\frac{\partial
\varepsilon(s,t)}{\partial s}\dd t =\\
\sum_{i=1}^m \tr\left\{\rho(0)\frac{\dd Q^{\dag}(T,s)}{\dd s} \left(\sum_k c_{ki}\Theta_i\right)
\frac{\dd Q(T,s)}{\dd s}\right\}~\textbf{e}_i.
\end{multline}
Based on eqn (\ref{grad1}), we have
$$\frac{\delta \textbf{v}_i(s)}{\delta \varepsilon(s,t)} =  \frac{1}{\i\hbar}\tr\left(
\left[\sum_k c_{ki}\Theta_{i}(T),\rho(0)\right]\mu(t)\right) $$
for the gradient of each of the observable expectation values
$\langle \Theta_i \rangle$. Following the above derivation, we can
convert this generalized differential equation into a vector
integral equation:
\begin{multline}
\sum_{i=1}^m\int_0^T \frac{1}{\i\hbar}\tr\left(
\left[\sum_k
c_{ki}\Theta_{i}(T),\rho(0)\right]\mu(t)\right)\textbf{e}_i~
\frac{\partial \varepsilon(s,t)}{\partial s}\dd t  \\
= \sum_{i=1}^m \tr\left\{\rho(0) \frac{\dd Q^{\dag}(T,s)}{\dd s} \left(\sum_k
c_{ki}\Theta_i\right) \frac{\dd Q(T,s)}{\dd
s}\right\}~\textbf{e}_i.
\end{multline}
Denoting the vector observable track of interest by $\textbf{w}(s)$, i.e.,
$$\textbf{w}(s) \equiv \sum_{i=1}^m \tr\left\{\rho(0) Q^{\dag}(T,s)
\left(\sum_k c_{ki}\Theta_i\right)
 Q(T,s)\right\}~\textbf{e}_i,$$
and expanding $\frac{\partial \varepsilon(s,t)}{\partial s}$ on the
basis of orthogonal observables, $$\frac{\partial
\varepsilon(s,t)}{\partial s} = \sum_{i=1}^m x_i \frac{\delta
\textbf{v}_i(T,s)}{\delta \varepsilon(s,t)},$$ we have $$\int_0^T
\left(\frac{\delta \textbf{v}(T,s)}{\delta
\varepsilon(s,t)}\right)^T ~\textbf{x}\cdot\frac{\delta
\textbf{v}(T,s)}{\delta \varepsilon(s,t)} \dd t= \frac{\dd
\textbf{w}(s)}{\dd s},$$ or equivalently,
$$\sum_{j=1}^m \int_0^T\frac{\delta \textbf{v}_i(T,s)}{\delta \varepsilon(s,t)}~~\textbf{x}_j\frac{\delta \textbf{v}_j(T,s)}{\delta \varepsilon(s,t)}\dd t= \frac{\dd \textbf{w}(s)}{\dd s}.$$
Defining the correlation matrix in this case as
$$\Gamma_{ij}(s) \equiv \int_0^T \frac{\delta \textbf{v}_i(T,s)}{\delta \varepsilon(s,t)}\frac{\delta \textbf{v}_j(T,s)}{\delta \varepsilon(s,t)}\dd t,$$
we obtain the following nonsingular algebraic differential equation
for the algorithmic step in the control field:
\begin{equation}\label{vectrack}
\frac {\partial E}{\partial s} = f_s(t) + \left[\frac{\dd \textbf{w}}{\dd s}-\textbf{a}(s) \right]^T\Gamma^{-1}\frac{\delta \textbf{v}(T,s)}{\delta \varepsilon(s,t)}
\end{equation}
where $f_s(t)$ is again a free function and we have defined the
vector function $\textbf{a}(s)$ by analogy to $\alpha(s)$ above:
$$\textbf{a}(s) \equiv \int_0^T\frac{\delta \textbf{v}(T,s)}{\delta \varepsilon(s,t)}f_s(t) \dd t.$$
The advantage of orthogonal observable expectation value tracking,
compared to unitary matrix tracking, is that the likelihood of the
matrix $\Gamma$ being ill-conditioned - even at abnormal extremal
control fields $\varepsilon(t)$, where $\textmd{G}$ is singular - diminishes
rapidly with $N^2-m$, where $m$ is the number of orthogonal
observable operators employed.

In the special case where only the observable of interest $\Theta_1$
is measured at each algorithmic step, this equation reduces to:
\begin{equation}\label{scalarflow}
\frac {\partial \varepsilon(s,t)}{\partial s} = f(s,t)+ \frac{\frac{\dd
P}{\dd s}
-\int_0^T a_0(s,t,T)f(s,t) \dd t}{\gamma(s)}a_0(s,t),
\end{equation}
where $P(s)$ is the desired track for $\langle \Theta_1(T) \rangle$, $a_0(s,t,T)\equiv-\frac{1}{i\hbar}\tr\left(\rho(0)
\big[U^{\dag}(T,0)\Theta_1U(T,0),U^{\dag}(t,0)\mu(s)U(t,0)\big]\right)$, and $\gamma(s)
\equiv \int_0^T \left[a_0(s,t,T)\right]^2\dd t$. Here, it is of
course not necessary to carry out any observable operator
orthogonalization.


Of course, measuring the expectation values (or gradients) of two or
more observable operators is more expensive than following the
gradient of a single observable. However, note that the gradients
$\frac{\delta \langle \Theta_1(T) \rangle}{\delta \varepsilon(t)}$
and $\frac{\delta \langle \Theta_2(T) \rangle}{\delta
\varepsilon(t)}$ of multiple observables are closely related since
$$\frac{\delta  \langle \Theta_1 \rangle}{\delta \varepsilon(t)} =
-\frac{i}{\hbar}\tr\{\left[U^{\dag}(T)\Theta_1U(T),\mu(t)\right]\rho(0)\}$$
while
$$\frac{\delta  \langle \Theta_2 \rangle}{\delta \varepsilon(t)} =
-\frac{i}{\hbar}\tr\{\left[U^{\dag}(T)\Theta_2U(T),\mu(t)\right]\rho(0)\}.$$
As such, the information gathered through the estimation of the
gradient of $\langle \Theta_1(T) \rangle$ can be used to "inform"
the estimation of $\langle \Theta_1(T) \rangle$. In particular,
although the norms of these two gradients differ, their
time-dependencies - i.e., $\frac{\delta \langle \Theta_1(T)
\rangle}{\delta \varepsilon(t_1)}/\frac{\delta \langle
\Theta_1(T) \rangle}{\delta \varepsilon(t_2)}$ are identical.
Hence, only one high-dimensional gradient estimation needs to be
carried out.

The above algorithm can be applied to follow an arbitrary set of
observable expectation value tracks $\{\langle \Theta_i(s)
\rangle\}$. Here, we are interested in following the observable
tracks that correspond to the shortest path between $U_0$ and $W$ on
the domain of unitary propagators, namely the geodesic path
$U(s)=U(0)\exp(iAs)$ with $A = -i\log(W^{\dag}U(s))$. As mentioned,
the matrix $W$ can be determined numerically if $\rho(0)$ and
$\Theta$ are known, for minimal computational cost. As shown by
Hsieh et al. \cite{Mike2006a}, there exists a continuous submanifold
of unitary matrices $W$ that solve the observable maximization
problem; if we denote the Hilbert space dimension by $N$, the
dimension of this submanifold ranges from $N$ in the case that
$\rho(0)$ and $\Theta$ are full rank nondegenerate matrices to
$N^2-2N+2$ in the case that $\rho$ and $\Theta$ are both pure state
projectors (see section \ref{phase}).

The dimension of the subspace $M_T$ of $\U(N)$ that is consistent
with the observed track $\frac{\dd \textbf{v}(T,s)}{\dd s}$ displays
a complicated dependence on the eigenvalue spectra of $\rho(0)$ and
$\{\Theta_i\}$. We demonstrate this explicitly for the case of
single observable tracking. In this case,
\begin{multline}
M_T \equiv \{V(s) \mid
\tr\big(V(s)^{\dag}\rho(0) V(s) \Theta\big) = \\
\tr\big(U(s)^{\dag}\rho(0)U(s) \Theta \big)=\langle \Theta(s)
\rangle\}
\end{multline}
where $U(s) = \exp{(i\log{(W^{\dag}U_0)}s)}$. As a first
step, we must characterize the degenerate subset $M(s)$ of unitary
matrices that are compatible with a given observable expectation
value $\langle \Theta \rangle$, as a function of the eigenvalue
spectra of $\rho(0)$ and $\Theta$. Let $\rho(0) = Q^{\dag}\epsilon
Q$ and $\Theta = R^{\dag}\lambda R$, where
$\epsilon_1,\epsilon_2,...$ and $\lambda_1,...,\lambda_2,...$ are
the eigenvalues of $\rho(0)$ and $\Theta$ with associated unitary
diagonalization transformations $Q$ and $R$ respectively. Then the
observable expectation value corresponding to a given unitary
propagator can be written
\begin{multline}
J(U)=\tr(U^{\dag}R\hat \rho(0)R^{\dag}US\hat \Theta S^{\dag}) \\
= \tr\left[(R^{\dag}US)^{\dag}\hat \rho(R^{\dag}US)\hat \Theta\right]=\tr(\hat U^{\dag}\rho(0)\hat U \hat \Theta)
\end{multline}
where the isomorphism $\hat U = R^{\dag}US$ also runs over $U(N)$,
and $U \equiv U(s)$. Therefore, without loss of generality, we can
always assume that both $\rho(0)$ and $\Theta$ are in diagonal form,
and determine $M(s)$ in terms of $\hat U(s)$, instead of $U(s)$.
Denote by $\U(\textbf{n})$ the product group $\U(n_1) \times \cdots
\times \U(n_r)$, where $\U(n_1)$ is the unitary group acting on the
$n_i$-dimensional degenerate subspace corresponding to $\lambda_i$,
and define $\U(\textbf{m})=\U(m_1)\times \cdots \U(m_s)$ in the same
manner. Then any transformation $\hat U \rightarrow Q \hat U
T^{\dag}$, where $Q \in \U(\textbf{n})$ and $T \in \U(\textbf{m})$,
leaves $J$ invariant:
\begin{eqnarray*}
\tr(T \hat U^{\dag}Q^{\dag}\hat \rho(0)Q \hat U T^{\dag} \hat \Theta) &=& \tr(\hat U^{\dag}Q^{\dag}\hat \rho(0)Q \hat U T^{\dag} \hat \Theta T)\\
&=& J(\hat U).
\end{eqnarray*}
This can be seen by observing that since $\hat \rho(0)$ is diagonal,
any unitary transformation $\hat \rho(0)\rightarrow Q^{\dag}\hat
\rho(0) Q = \hat \rho(0) Q^{\dag}Q = \hat \rho(0)$, if the unitary
blocks of $Q$ are aligned with the degeneracies of $\rho(0)$. By the
cyclic invariance of the trace, we also have $\hat \Theta
\rightarrow T^{\dag}\hat \Theta T = \hat \Theta T^{\dag}T=\hat
\Theta.$ Thus, the degenerate manifold can be written $M(s) =
\U(\textbf{n})\hat U(s)\U(\textbf{m})$. Hence, the entire subspace
of $\U(N)$ that is accessible to the system propagator during global
observable tracking is
$$M_T = \bigcup_{0\leq s \leq 1}\U(\textbf{n})\hat U(s)\U(\textbf{m}).$$

The manifold $M(s)$  can be expressed as the quotient set $$M(s)=\frac{\U(\textbf{n})\times \U(\textbf{m})}{\U(\textbf{m}) \cap \hat U^{\dag}(s) \U(\textbf{n}) \hat U(s)},$$
which can be seen as follows. Define $F_{H}(P,Q): H \rightarrow P H Q$, where $H \in \U(N)$ and $(P,Q) \in \U(\textbf{n})\times \U(\textbf{m})$.
Let $stab(H)$ denote the stabilizer of $H$ in $\U(\textbf{n})\times \U(\textbf{m})$, i.e. the set of
matrix pairs $(X,Y) \in \U(\textbf{n}) \times \U(\textbf{m})$ such that $F_H(X,Y) = X H Y = H$. The stabilizer
characterizes the set of points that are equivalent with $H$, hence the manifold $M$ can be identified
as the quotient set of $\U(\textbf{n})\times \U(\textbf{m})$ divided by $stab(H)$. We can specify
the stabilizer as follows. First, from $Y=H^{\dag}U^{\dag}H$, we see that $H$ transforms $U \in \U(\textbf{n})$
into $\U(\textbf{m})$. Hence $Y \in \U(\textbf{m}) \cap H^{\dag}\U(\textbf{n})H$. Conversely, for any
$Y \in \U(\textbf{m}) \cap H^{\dag}\U(\textbf{n}) H$, the pair $(Y^{\dag}HY,Y)$ must be a member of $stab(H)$.
Hence, the stabilizer is isomorphic to $\U(\textbf{m}) \cap H^{\dag}\U(\textbf{n}) H$. In the present case
where $H = U(s)$, we have
\begin{multline}
stab(\hat U(s)) = \{(\hat U(s) Y^{\dag}\hat U^{\dag}(s),Y): \\
Y \in \U(\textbf{m}) \cap \hat U^{\dag}(s) \U(\textbf{n}) \hat U(s)\}.
\end{multline}
Thus the dimension of the degenerate manifold $M(s)$ is $$D_0(M(s)) =\textmd{dim}\: \U(\textbf{n}) +\textmd{dim}\: \U(\textbf{m})-\textmd{dim}\; stab(\hat U(s)).$$

The dimension of this subspace cannot be specified in a simple form
for arbitrary $U(s) \in \U(N)$, since it is governed by the
dimension of the stabilizer. We note a couple of special cases. If
$\hat U(s) = R^{\dag} U(s) S$ contains unitary subblocks that fall
within the overlapping unitary subblocks in $U(\textbf{n})$ and
$U(\textbf{m})$,the dimension of the stabilizer is at least as large
as that of the subblocks. If $\hat U(s) = R^{\dag} U(s) S$ is a
permutation matrix $\Pi$ or a product of a permutation matrix with a
matrix of the aforementioned type, the permutation matrix can act to
rearrange the subblocks of $U(\textbf{m})$ so that they overlap with
those of $U(\textbf{n})$ and thereby increase the dimension of the
stabilizer. The dimension of the subspace (subgroup) of $\U(N)$
composed of such matrices $U(s)$ can be shown to increase very
rapidly with increasing degeneracies $m$ and $n$ in $\rho(0)$ and
$\Theta$, respectively.

The matrix $\hat U(s)$ can only rearrange or diminish the size of
existing subgroups $\U(n_i)$ within $\U(\textbf{n})$, but cannot
create larger subgroups. Thus, we can establish a maximal dimension
for $stab(\hat U(s))$  for any given $\hat U(s)$ as that which
maximizes the overlap between the respective subgroups $\U(m_j)$ and
$\U(n_i)$. This bound is achieved when the conjugation action
$U(s)^{\dag}\U(\textbf{n}) U(s)$ of $\hat U(s)$ is equivalent to the
action $\Pi(s)^{\dag}\U(\textbf{n}) \Pi$ of the permutation matrix
$\Pi$ that rearranges the subblocks such that they display maximal
overlap. Therefore, for fixed $\rho(0)$, $\Theta$, the dimension of
the manifold can range from the maximal value of $\sum_in_i^2 +
\sum_jm_j^2 - N$ down to $\sum_{i=1}^r n_i^2 + \sum_{j=1}^s m_j^2 -
\sum_{1\leq i \leq r, 1 \leq j \leq s} k_{ij}^2$ in the case that
the conjugation action of $\hat U(s)$ satisfies the above condition,
where $k_{ij}$ denotes the number of positions in the diagonal where
the eigenvalues $\lambda_i$ and $\epsilon_j$ appear simultaneously
after imposition of the permutation matrix $\Pi$.

Thus, we see that the size of the set of unitary matrices $V$
producing the same observable expectation value $\langle
\Theta\rangle$ will change along the trajectory $U(s)$, based on the
extent to which the latter reorients the eigenvalues of $\rho(0)$
and $\Theta$ such that degeneracies coincide. In particular, the volume of
the subspace $M_T$ of $\U(N)$ that is consistent with a track
$\langle \Theta(s) \rangle$ derived from a geodesic between $U_0$
and $W$ will display a strong dependence on the choice of $U_0$ and
$W$, and will change depending on the matrix $W$ that is chosen from
the degenerate submanifold of unitary matrices that solves the
observable maximization problem.

Note that the coefficient $a_0$ in the (single) observable
expectation value tracking differential equation is in fact equal to
the gradient on the domain $\varepsilon(t)$, equation (\ref{grad1}). Recall
that the gradient flow (\ref{Egrad}) is defined by the differential
equation $\frac{\dd E}{\dd s} = \frac{\delta \langle \Theta
\rangle}{\delta E}$. Since the coefficients of $a_0$ are scalars, we
see that the algorithmic path for scalar tracking can be expanded on
a basis whose dimension is identical to that of the gradient basis
(\ref{gradbasis}), as expected. We will analyze the dependence of
the dimension of this basis on the eigenvalue spectra of $\rho(0)$
and $\Theta$ in section \ref{integration}.

As a function of the algorithmic step $s$, the coefficient
$\frac{\int_0^T a_0(s,t,T)f(s,t) \dd t}{\gamma(s)}$ in eqn (\ref{scalarflow}) will adjust
the step direction so that unitary matrices $V(s)$ at each step are
constrained within the subspace $M(s)$. According to the above
analysis, this dimension of this subspace will scale more steeply
with increasing degeneracies in $\rho(0)$ and $\Theta$. The maximal
dimension of $M(s)$ ranges from $N^2-N$ for the problem where $\rho$
and $\Theta$ are both pure state projectors, to $N$ for the case
where $\rho$ and $\Theta$ are full rank with completely
nondegenerate eigenvalues. Even in the former case, this represents
an advantage over the $\varepsilon(t)$-gradient flow, which is free to explore
the full $N^2$-dimensional space of dynamical propagators in
$\U(N)$.

The above analysis assumes that the initial density matrix $\rho(0)$
is known to arbitrary precision. This information is, of course, not
required for $\varepsilon(t)$-gradient based optimization, but it is readily
acquired in the case that the initial state is at thermal
equilibrium. For orthogonal observable tracking, the cost of partial
quantum state reconstruction of $\rho(T)$ at each algorithmic step
must be weighed against the increase in efficiency obtained by
virtue of following a globally optimal path (section \ref{oce}).

\section{Error correction and fluence minimization}\label{errorfluence}

\subsubsection{Error correction}

In attempting to track paths on $\U(N)$, errors will inevitably occur for two reasons. First, the
algorithmic step on $\U(N)$ will be a linear approximation to the
true increment $\delta U(T)$ due to discretization error; this error
will increase as a function of the curvature of the integrated flow
trajectory at algorithmic time $s$. Second, the D-MORPH integral
equation is formulated in terms of only the first-order functional
derivative $\frac {\delta U(T)}{\delta \varepsilon(t)}$ (or $\frac {\delta \textbf{v}(T)}{\delta \varepsilon(t)}$, $\frac {\delta \langle \Theta(T) \rangle}{\delta \varepsilon(t)}$ for
orthogonal observable and single observable tracking, respectively); the error incurred by
neglecting higher order terms in the Taylor expansion will depend on
the system Hamiltonian.

In our numerical simulations, we apply error-correction methods to account for these deviations
from the track of interest. (These methods can in principle also be implemented in an experimental setting.) For unitary matrix tracking, we correct for these inaccuracies by
following the (minimal-length) geodesic from the real point
$U(T,s_k)$ to the track point $Q(s_k)$. This correction can be
implemented by incorporating the function $C(s_k)=-\frac{i}{s_{k+1}-s_k}\log(Q^{\dag}(s_k)U(T, s_k))$ into the matrix differential
equation for the algorithmic time step:
$$\frac{\partial E}{\partial
s} =  f(s,t) + \Big( v(C(s)) + v(\Delta) -\alpha
\Big)^T\textmd{G}^{-1}v({\mu(t)})$$
In a more efficient approach, we combine error correction and the
next gradient step in one iteration \cite{Dominy2007}. In this case, we define
$\Delta(s_k)=-\frac{i}{s_{k+1}-s_k}\log
\left(Q^{\dag}(s_{k+1})U(T,s_k)\right)$ and use $$\frac{\partial E}{\partial s} = f(s,t) + \Big(v(\Delta) -\alpha
\Big)^T\textmd{G}^{-1}v({\mu(t)}).$$

For orthogonal observable expectation value tracking, the vector
space within which $\textbf{v}(s)$ resides is not a Lie group, and
consequently it is not as straightforward to apply error correction
algorithms that exploit the curved geometry of the manifold. We
therefore choose the error correction term to be a simple scalar
multiple of the difference between the current values of the observable vector
and its target value, i.e. $\beta\left[\textbf{w}(s)-\textbf{v}(s)\right]$, such that the
tracking differential equation becomes
\begin{multline}
\frac {\partial E}{\partial s} = f(s,t) ~+ \\
\left[\beta\left(\textbf{w}(s)-\textbf{v}(s)\right)+\frac{\dd \textbf{w}}{\dd s}- \textbf{a}(s)\right]^T\Gamma^{-1}\frac{\delta \textbf{v}(T,s)}{\delta \varepsilon(s,t)}.
\end{multline}
For the special case of single observable tracking, this reduces to
\begin{multline}
\frac {\partial \varepsilon(s,t)}{\partial s} = f(s,t)~+ \\
\frac{\beta(P(s)-\langle \Theta(s) \rangle + \frac{\dd P(s)}{\dd s} -\int_0^T a_0(s,t,T)f(s,t) \dd t}{\gamma(s)}a_0(s,t).
\end{multline}

\subsubsection{Fluence minimization}

Clearly, the above analysis does not take into account the common
physical constraint of penalties on the total field fluence. The
effect of the fluence penalty in this scenario is then to decrease
the degeneracy in the solutions to the above system of equations for
$\frac {\partial \varepsilon(s,t)}{\partial s}$. This is accomplished by
choosing the free function $f(s,t)$ in either the unitary or orthogonal observable
tracking differential equations to be an explicit function
of the electric field. It can be shown \cite{Rothman2005} that the choice:
$$f(s,t) = - \frac{1}{\Delta s} \varepsilon(s,t) W(t),$$
where $W(t)$ is an arbitrary weight function and the $\Delta s$ term
controls numerical instabilities, will determine the $\frac
{\partial \varepsilon(s,t)}{\partial s}$ at each algorithmic time step $s$
that minimizes fluence.

\section{Integration of quantum observable expectation value gradient
flows} \label{integration}

We have seen (eqn \ref{Gmat}) that in general, the projected path in
$\U(N)$ that originates from following the local $\varepsilon(t)$-gradient
depends on the Hamiltonian of the system through the matrix
$\textmd{G}$. Nonetheless, there is still a Hamiltonian independent
component to the $\varepsilon(t)$-gradient, which corresponds to the gradient
on the domain $\U(N)$. Thus, it is of interest to gain some
understanding of the behavior of this $U$-gradient flow - whether it
follows a direct path toward the target unitary propagator, or
whether it biases the $\varepsilon(t)$-gradient flow to follow indirect paths
in $\U(N)$. Such an analysis may shed light on the comparative
optimization efficiencies of the gradient compared to the tracking
algorithms described in the previous sections.

It can be shown that the $U(T)$ and $\varepsilon(t)$ gradient flows evolve
locally on subspaces of the same dimension, and that this dimension
changes predictably as a function of the eigenvalue spectra of
$\rho(0)$ and $\Theta$. These gradient flows evolve on a subspace of
the homogeneous space of $\U(N)$ whose dimension is given by the
spectrum of the initial density matrix $\rho(0)$, necessitating the
use of a distinct coordinate basis to express the integrated
gradient flow trajectories for different classes of $\rho(0)$ that
depend on the latter's number of nonzero and degenerate eigenvalues
\cite{HoRab2007b}. Specifically, let $\rho(0)$ consist of $r$ subsets
of degenerate eigenvalues  $p_1, \cdots, p_r$, with multiplicities
$n_1,\cdots, n_r$. Writing $\rho(0)= \sum_{i=1}^n p_i|i\rangle
\langle i|$, we have
\begin{multline}
\frac{\delta \Phi}{\delta \varepsilon(t) } = a_0(t,T) = \\
\frac{i}{\hbar} \sum_{i=1}^n p_i \sum_{j=1}^N \left[ \langle i|\Theta(T)|j \rangle
\langle j|\mu(t)|i \rangle - \langle i|\mu(t)|j \rangle \langle
j|\Theta(T)|i\rangle \right]
\end{multline}
Defining $s_k \equiv \sum_{i=1}^{k-1}n_i, \quad k=2,\cdots, r+1$ this can be written
\begin{widetext}
\begin{eqnarray}
\frac{\delta \Phi}{\delta \varepsilon(t)}&=&\frac{i}{\hbar} \sum_{k=1}^r p_k
\sum_{i=s_k+1}^{s_{k+1}} \left[\sum_{j=1}^{s_k} +
\sum_{j=s_k+1}^{s_{k+1}} + \sum_{j=s_{k+1}+1}^N \right] \{\langle
i|\Theta(T)|j \rangle \langle j|\mu(t)|i \rangle -\langle i|\mu(t)|j
\rangle \langle j|\Theta(T)|i \rangle \}\\
&=&\frac{i}{\hbar} \sum_{k=1}^r p_k \sum_{i=s_k+1}^{s_{k+1}}\left[\sum_{j=1}^{s_k} +
\sum_{j=s_{k+1}+1}^N \right] \{\langle i|\Theta(T)|j\rangle \langle
j|\mu(t)|i\rangle - \langle i|\mu(t)|j \rangle \langle j|\Theta(T)|i
\rangle \}
\end{eqnarray}
\end{widetext}
where the second equality follows from the fact that the terms corresponding to $\sum_{j=s_k+1}^{s_{k+1}}$ (i.e., those arising from the same
degenerate eigenvalue of $\rho(0))$) are zero. This indicates that the dimension of the subspace of the
space of skew-Hermitian matrices upon which the gradient flow
evolves is
$$D=N^2-(N-n)^2-\sum_{i=1}^r n_i^2 = n(2N-n)-\sum_{i=1}^r n_i^2.$$
This is the dimension of a compact polytope $P$ which is the convex
hull of the equilibria of the gradient vector field. The gradient
flow involves on the interior of this polytope \cite{Bloch1992}.

Both the $\varepsilon(t)$-gradient flow (\ref{Egrad}) and observable tracking
(\ref{scalarflow}) can be expanded on this basis corresponding to
$\frac{\delta \Phi}{\delta \varepsilon(t) }$. It can be shown (see below) that
in the case of the gradient, the increased dimension of this basis
set for increasing nondegeneracies in $\rho(0)$, $\Theta$ generally
results in increased unitary pathlengths; since the entire unitary
group is free for exploration, the increased number of locally
accessible directions results in the path meandering to more distant
regions of the search space. By contrast, in the case of observable
tracking, the increased number of locally accessible directions for
greater nondegeneracies in $\rho(0)$, $\Theta$ are coupled with a
decrease in the dimension of the globally accessible search space in
$\U(N)$; the greater local freedom is used to follow the target
unitary tracks of interest.

In order to shed light on the origin of the aforementioned behavior
of the $\varepsilon(t)$-gradient, we consider the global paths followed by
$U$-gradient flow of $\Phi$. In a useful analogy, the control
optimization process can itself be treated as a dynamical system.
The critical manifolds of the objective function then correspond to
equilibria of the dynamical system, and the gradient flow
trajectories to its phase trajectories. Within this analogy, the
gradient flow of $\Phi$ on $\U(N)$ can be shown to represent the
equations of motion of an integrable dynamical system. The
expression (\ref{Uflow}) for the gradient flow of $\Phi$ above is
cubic in $U$. However, through the change of variables
$U(s,T)\rightarrow \rho(s,T) = U(s,T)\rho(0,0)U^{\dag}(s,T)$ we can reexpress it as a
quadratic function:
\begin{multline}
\dot \rho(s,T) = \\
-U(s,T)\rho(0,0)\dot U^{\dag}(s,T) -\dot U(s,T) \rho(0,0)U^{\dag}(s,T) \\
 = \rho^2(s,T) \Theta - 2\rho(s,T) \Theta \rho(s,T) + \Theta \rho^2(s,T)\\
 =\left[\rho(s,T),[\rho(s,T),\Theta ] \right]
\end{multline}
where $s$ denotes the algorithmic time variable of the gradient
flow in $\U(N)$ and the dot denotes the $s$-derivative. This quadratic expression for the gradient flow
is in so-called double bracket form
\cite{Brockett1991,Bloch1992,Helmke1994}. The set of all
$U(s,T)\rho(0,0)U^{\dag}(s,T)$ is a homogeneous space $M(\rho)$ for the Lie
group $\U(N)$, namely the space of all Hermitian matrices with
eigenvalues fixed to those of $\rho(0,0)$. Maximizing $\Phi(U)$ over
$\U(N)$ is equivalent to minimizing the least squares distance $\|
\Theta-U \rho U^{\dag} \|^2$  of $\Theta$ to $U\rho U^{\dag}
\in M(\rho)$:
$$\| \Theta-U\rho U^{\dag} \|^2 = \| \Theta \|^2 - 2\Phi(U) +
\| \rho \|^2.$$

Here, we provide an explicit formula pertaining to the analytical
solution for the above gradient flow for what is perhaps the most
common objective in quantum optimal control theory and experiments,
namely the maximization of an arbitrary observable starting from a
pure state. In particular, this includes the special case of
maximizing the transition probability $P_{if}$ between given initial
and final pure states $|i\rangle$ and $|f\rangle$. Whenever $\rho(t=0)$
is a pure state, $n=r=n_1=1$, and $D=2N-2$, and the convex hull of
the critical points of the vector field is a $(N-1)$-dimensional
simplex. The gradient flow evolves in the interior of this simplex.
The analytical solution for the fully general case of a mixed state
$\rho(0)$ and nondegenerate $\Theta$ is more complicated and
presented in another work of the authors.

Since the objective function is symmetric with respect to $\rho(0)$ and $\Theta$, this formulation
applies if either $\rho(0)$ or $\Theta$ is a pure state projection
operator, i.e., if at least one of them can be diagonalized by an appropriate
change of basis to matrices that have only one nonzero diagonal
element, corresponding to $|i\rangle \langle i|$ or $|f \rangle
\langle f|$, respectively.  The other operator can have an arbitrary spectrum.
The same integrated gradient flow thus applies to the problem of
maximizing the transition probability between any generic mixed initial state to any pure state.

Under these conditions, we can execute a change of variables such
that the double bracket flow, which evolves on the $\frac{1}{2}
m(m+1)$-dimensional vector space of Hermitian matrices $\rho(0)$ is
mapped to a flow on the m-dimensional Hilbert space. Letting
$|\psi(s) \rangle =U(s)|i\rangle$, the double bracket flow can be
written:
\begin{eqnarray*}
|\dot \psi(s)\rangle &=&\dot U(s)|i\rangle\\
&=&\left[\Theta U(s)|i\rangle -U(s)|i\rangle\langle i|U^\dagger(s)\Theta U(s)\right]|i\rangle \\
&=& \left[ \Theta - \langle \psi(s)|\Theta|\psi(s) \rangle I\right] ~ |\psi(s)
\rangle.
\end{eqnarray*}
If we define $x(s) \equiv (|c_1(s)|^2,\cdots,|c_N(s)|^2)$, where $c_1(s),\cdots,c_N(s)$ are
the coordinates of $|\psi(s)\rangle$ under the basis that
diagonalizes $\Theta$; it can be verified that the integrated
gradient flow can be written:
\begin{eqnarray}
x(s) &=& \frac{e^{2s\Theta}\cdot(|c_1(0)|^2,\cdots,|c_N(0)|^2)}{\sum_{i=1}^N|c_i(0)|^2e^{2s\lambda_i}} \\
&=&\
\frac{e^{2s\lambda_1}|c_1(0)|^2,\cdots,e^{2s\lambda_N}|c_N(0)|^2}{\sum_{i=1}^N|c_i(0)|^2e^{2s\lambda_i}}
\end{eqnarray}
where $\lambda_1,\cdots, \lambda_N$ denote the eigenvalues of
$\Theta$.

In the case that $\Theta$ has only one nonzero eigenvalue (pure
state), this becomes:
\begin{widetext}
\begin{eqnarray*}
x(s)&=& \left(\frac{|c_1(0)|^2}{\sum_{i\neq
j}^m|c_i(0)|^2+e^{2s\lambda_j}|c_j(0)|^2},\cdots,\frac{e^{2s\lambda_j}|c_j(0)|^2}{\sum_{i\neq
j}^m|c_i(0)|^2+e^{2s\lambda_j}|c_j(0)|^2},\cdots \right)\\
&=&
\left(\frac{|c_1(0)|^2}{1+(e^{2s\lambda_j}-1)|c_j(0)|^2},\cdots,\frac{e^{2s\lambda_j}|c_j(0)|^2}{1+(e^{2s\lambda_j}-1)|c_j(0)|^2},\cdots
\right).\end{eqnarray*}
\end{widetext}

In contrast to quantum time evolution of the state vector (which
resides on the Hilbert sphere $S^{m-1}$, the matrix $e^{2s\Theta}$
translates the vector $x$, which resides on the
$(m-1)$-dimensional simplex, through algorithmic time. Since
coherent quantum time evolution cannot change the eigenvalues of
the density matrix $\rho$, the optimization of controls simply
reorders these eigenvalues. Hence the gradient flow is said to be
an "isospectral" flow. We are primarily interested in the gradient
flow on the domain of unitary propagators of the quantum dynamics,
$U(T)$. In general, there exists a one-to-many map between
$\rho(T)$ and $U(T)$. This attests to the existence of a
multiplicity of paths through the search space of the quantum
optimal control problem that will maximize the objective
functional in equivalent dynamical time. Although the analytical solution we have presented is framed on the homogeneous of
$\U(N)$ rather than on the Lie group itself, we will see in the next section
that generic properties of the behavior of the flow trajectories, in particular their phase behavior
with respect to local critical points, are identical on these domains.

The above gradient flow for $|\psi\rangle $ only applies when
$\Theta$ has no degeneracy in its eigenvalues, including zero
eigenvalues. If the maximum eigenvalue of $\Theta$ is degenerate
with multiplicity $k$, such that $c_1=c_2=\cdots=c_k$, and $c_k >
c_j, \quad j=k+1,\cdots,n$, then the dynamics converges to the
point $\frac{1}{k}(1,\cdots,1,0,\cdots,0).$

A remarkable feature of the gradient flow for objective function
$\Phi$ is that it is a Hamiltonian flow \cite{Faybu1991,
Bloch1990, Bloch1995}, for general $\rho(0)$ and $\Theta$.
The eigenvalues can be viewed as the analog of
the momenta in the corresponding Hamiltonian system. The "isospectral" character of the flow indicates these momenta are conserved.
An alternative proof of the integrability of the flow for $\Phi$ is based on demonstrating
that in N dimensions, the flow has N integrals of the motion that
are in involution, which is the classical definition of complete
integrability for a Hamiltonian system. From the point of view of
the modern theory of integrable systems, the double bracket flow
can be shown to represent a type of Lax pair, a general form that
can be adopted by all completely integrable Hamiltonian systems
\cite{Babelon2003}.

\section{Phase behavior of quantum observable maximization gradient flows}\label{phase}

The integrated flow trajectories provided above for the gradient of $\Phi$ on
the domain of unitary propagators can be used to provide insight into global
behavior of the $\varepsilon(t)$-gradient flow.
Just as the global trajectory of the unitary geodesic flow influences, but does
not directly determine the unitary path followed by global observable tracking,
the integrated $U(T)$-flow trajectories influence but do not completely determine
the behavior of the $\varepsilon(t)$-gradient flow. A useful metric for assessing the
phase behavior of the $U(T)$-gradient flows is the distance of the search trajectory
from the critical manifolds of the objective as a function of algorithmic time.

The distance of the search trajectory to the global optimum of the objective can be expressed:

$$\|x(s)-e_{i*}\|^2 =
\|x(s)\|^2-2\frac{\langle e^{2s\Theta}x(0),e_{i*}\rangle }{\langle
e^{2s\Theta}x(0)\rangle }+1$$
The equilibrium points of this flow are the critical points of the objective function. Hsieh et al. \cite{Mike2006a} showed that the critical manifolds of $\Phi$
satisfy the condition $$\frac{\dd J}{\dd s}= i\tr\left(A\left[\Theta,U^{\dag}\rho(0)U\right]\right)=0.$$
The critical manifolds are then given by matrices of the form
$$\hat U_l = QP_lR^{\dag},$$ where $P_l, \quad l=1,\cdots,N!$ is an $N$-fold permutation matrix whose
nonzero entries are complex numbers
$\exp(i\phi_1),...,\exp(i\phi_N)$ of unit modulus, and
$\rho=Q^{\dag}\epsilon Q$ and $\Theta = R^{\dag}\lambda R$. The
critical manifolds have a similar topology to the manifolds $M(s)$
shown in section \ref{orthog} to map to a given observable
expectation value. For example, in the case that $\rho$ and $\Theta$
are fully nondegenerate, they are $N$-torii $T_l^N$. In this case,
the number of critical manifolds scales factorially with the Hilbert
space dimension $N$. In the case that either $\rho$ or $\Theta$ has
only one nonzero eigenvalue, while the other is full rank and
nondegenerate, the number of critical manifolds scales linearly with
$N$. (In the present case, assuming $\Theta$ is full-rank, the
number of these equilibrium points scales exponentially as $2^m$
with the Hilbert space dimension.) The optimal solution to the
search problem corresponds to the basis vector where $\Theta$ has
its maximal eigenvalue. In the case that the observable operator
$\Theta$ has only one nonzero eigenvalue $i$, there are only two
critical points, corresponding to $\pm e_i$.

The time derivative of the distance of the search trajectory to a critical manifold (framed on the homogeneous space)
is

\begin{equation}
\frac{\dd}{\dd t}\|x(s)-e_{i*}\|^2
=\frac{e^{2s\lambda_i*}x_{i*}(0)\sum_{j=1}^m(\lambda_{i*}-\lambda_j)e^{2s\lambda_j}x_j(0)}{\left[\sum_{j=1}^me^{2s\lambda_j}x_j(0)\right]^2}.
\end{equation}
Solving for the zeroes of this time derivative reveals that the
distance between the current point on the search trajectory and the
solution can alternately increase and decrease with time. The
distance of the gradient flow trajectory from the suboptimal
critical manifolds can alternately increase and decrease arbitrarily
depending on the spectral structure of $\Theta$. Thus the density
matrix does not always resemble the target observable operator to a
progressively greater extent during the algorithmic time evolution.

Qualitatively, the most obvious and important feature of these trajectories is that
the closer the initial condition is to a suboptimal critical
manifold, the greater the extent to which the gradient flow follows
the boundary of the simplex during its early time evolution. Away
from these initial conditions, the behavior of the gradient flow
trajectory is considerably more sensitive to the spectral structure
of the observable operator $\Theta$ than it is to the initial state
$\psi_0$. Indeed, it may be shown \cite{RajWu2007} that for operators $\Theta$
with two eigenvalues arbitrarily close to each other, the time
required for convergence to the global optimum increases without
bound, whereas this is not the case for initial states with $c_i$
arbitrarily close to 0.

Of course, the search trajectory slows down in the vicinity of the critical manifolds.
Note that at the critical manifolds themselves, the observable tracking
equations also encounter singularities, since $a_0 = 0$. In the vicinity
of the critical manifolds (the attracting regions) tracking may be less
accurate due to small values of $\gamma(s)$ in the denominator of eqn (\ref{scalarflow}), corresponding to
functions $a_0$ of small modulus.

As the nondegeneracies in $\rho(0)$ and $\Theta$ increase, the number
of attractors of the search trajectory increases, indicating an
increase in the unitary path length. Therefore, as mentioned above,
although the trajectory followed by the gradient locally accesses
the same number of directions as observable tracking, its global
path appears biased towards being longer, assuming the Hamiltonian
is fixed, especially for highly nondegenerate $\rho(0)$, $\Theta$; the
local steps in global geodesic observable tracking access the same
number of local directions but orient them towards unitary matrices
along a shorter path.

\section{Numerical implementation}\label{numerical}

Simulations comparing the efficiency of unitary or orthogonal observable tracking with
gradient-based optimal control algorithms will be presented in a separate paper.
However, we provide here a brief summary of numerical methods that can be used to implement the
various tracking algorithms described above.

For multiple observation-assisted tracking, a set of $n$ observable
operators $\Theta_1,\cdots, \Theta_n$ were either chosen randomly or
based on eigenvalue degeneracies. This (possibly
linearly dependent) set was orthogonalized by Gram-Schmidt
orthogonalization, resulting in a $m$ (where $m \leq N$) dimensional
basis set of observable operators. The $m$-dimensional vector $v(s)$
was constructed by tracing each of these observable operators with
the density matrix.

Numerical solution of the D-MORPH differential equations (\ref{utrack}, \ref{vectrack}, \ref{scalarflow})
was carried out as follows. The electric field $\varepsilon(s,t)$ was stored
as a $p\times q$ matrix, where $p$ and $q$ are the number of
discretization steps of the algorithmic time parameter $s$ and the
dynamical time $t$, respectively. For each algorithmic step $s_k$,
the field was represented as a $q$-vector for the purpose of
computations. Starting from an initial guess $\varepsilon(s_0,t)$ for the
control field, the Hamiltonian was integrated over the interval
$~[0,T~]$ by propagating the Schrodinger equation over each time
step $t_k -> t_{k+1}$, producing the local propagator
$U(t_{j+1},t_j) = \exp~[-iH(s_i,t_j)T/(q-1)~]$. For this purpose,
the propagation toolkit was used. Local propagators were precalculated via
diagonalization of the Hamiltonian matrix (at a cost of $N^3$), exponentiation of the
diagonal elements, and left/right multiplication of the resulting matrix by the
matrix of eigenvectors/transpose of the matrix of eigenvectors.
This approach is generally faster than
computing the matrix exponential directly. Alternatively, a
fourth-order Runge-Kutta integrator, can be
employed for the propagation, and is often used for density matrix
propagation.

The time propagators $U(t_j,0)=U(t_j,t_{j-1}),\cdots,U(t_1,t_0)$
computed in step 1 were then used to calculate the time-evolved
dipole operators $\mu(t_j)=U^{\dag}(t_j,0)\mu U(t_j,0)$, which can
be represented as a $q$-dimensional vector of $N\times N$ Hermitian
matrices. The $N^2 \times N^2$ matrix $\textmd{G}(s_k)$ and $N^2$ vector
$\alpha(s_k)$ (alternatively, the $m \times m$ matrix $\Gamma(s_k)$ and
$m$ vector $\textbf{a}(s_k))$) were then computed by time integration of the dipole
functions (and appropriate choice of function $f(s,t)$ )described
above. For tracking of unitary gradient flows, the next point
$Q(s_{k})$ on the target unitary track, necessary for the
implementation of error correction, was calculated in one of two
ways: i) numerically through $Q(s_{k})=Q(s_{k-1})*\exp(-i\Delta(s_k)
\dd s)$, computed using a matrix exponential routine (because of the
greater importance of speed vs accuracy for this step) or ii)
analytically, through the integrated flow equation above. If error
correction was employed, the matrix $C(s_k)=-\frac{i}{s_k - s_{k-1}}
\log (U^{\dag}(s_k)Q(s_k))$ for unitary error correction was
calculated by diagonalization of the unitary matrix followed by
calculation of the scalar logarithms of the diagonal elements, where
the logarithms are restricted to lie of the principal axis.

Next, the control field $\varepsilon(s_k,t)$ was updated to $\varepsilon(s_{k+1},t)$.
This step required the inversion of the $N^2 \times N^2$ matrix
$\textmd{G}(s_k)$ or $m \times m$ matrix $\Gamma(s_k)$, which was carried out using LU decomposition. The
quantities $\textmd{G}^{-1}(s_k)$, $\Delta(s_k)$, $\alpha(s_k)$
(alternatively, $\Gamma^{-1}(s_k), \frac{\textbf{v}(s)}{\dd s},
\textbf{a}(s_k)$, for scalar tracking) and $C(s_k)$ were used to compute
the $q$-dimensional vector $\frac{\partial \varepsilon(s_{k},t)}{\partial s}$.
One of two approaches was used to update the field: (i) a simple
linear propagation scheme, i.e. $\varepsilon(s_{k+1},t) = \varepsilon(s_{k},t) + \dd s ~
\frac{\partial \varepsilon(s_{k},t)}{\partial s}$, or (ii) a fourth-order
Runge-Kutta integrator. Because the accuracy of tracking depended
largely on the accuracy of this $s$-propagation step, and because
only one $s$-propagation was carried out for each set of $q$ time
propagations, a more accurate (but expensive) fifth-order
Runge-Kutta integrator was often used in this step. The updated
control field $\varepsilon(s_{k+1},t)$ was then again used to the propagate
the Schrodinger equation.

In the above numerical tracking scheme, the most computationally
intensive step is the propagation of the Schrodinger equation.
Although calculation of the matrix elements of $\textmd{G}$ for unitary tracking (respectively,
$\Gamma$ for orthogonal observable tracking) and
inversion of this matrix scale unfavorably with the
dimension of the quantum system, this scaling is polynomial. Since
remaining on the target $U$-track can play an important role in the
global convergence of the algorithm, especially where the local
gradient follows a circuitous route, the additional expense incurred
in accurate $s$-propagation may be well-warranted.

\section{Implications for optimal control experiments (OCE)}\label{oce}

The majority of optimal control experiments (OCE) have been
implemented with adaptive or genetic search algorithms, which
only require measurement of the expectation value of the observable for
a given control field. Recently, gradient-following algorithms have been implemented
in OCE studies, based on the observation that the control landscape is
devoid of suboptimal traps \cite{Roslund2007}. In computer simulations,
it is generally observed that gradient-based algorithms converge more
efficiently than genetic algorithms. However, more sophisticated local search algorithms,
such as the Krotov or iterative algorithms often used in OCT, are difficult if not impossible to
implement experimentally due to the extreme difficulty in measuring second
derivatives of the expectation value in the presence of noise and decoherence.
Therefore, global search algorithms which can be implemented on the
basis of gradient information alone, such as orthogonal observable tracking,
are particularly attractive candidates for improving the search efficiency of OCE.

Applying observable tracking experimentally requires a fairly simple
extension of gradient methods that have already been applied. The
gradient is determined through repeated measurements on identically
prepared systems, to account for the impact of noise. Instead of
following the path of steepest descent, the laser field is updated
in a direction that in the linear approximation would produce the
next observable track value $\Theta(s+\dd s)$. The assumption is
that because this observable step is consistent with a short step on
the domain of unitary propagators, the associated error in reaching
this expectation value will be smaller than that associated with
trying to move over the same fraction of the gradient flow
trajectory. Error correction can be implemented using a method
identical to that described above for numerical simulations.

In the simplest incarnation of orthogonal observation-assisted
quantum control, the full density matrix $\rho(0)$ is estimated at
the beginning of the control optimization. This need be done only
once and hence adds only a fixed overhead to the experimental effort
that does not add substantially to the scaling of algorithmic cost
with system dimension. At subsequent steps during the optimization,
the experimenter can (possibly adaptively) decide how many
orthogonal observations to make in order to estimate the final
density matrix $\rho(T)$, with the goal of keeping the unitary path
as close as possible to the desired geodesic. The number of distinct
observables that must be measured at each step (and hence roughly
the total number of measurements) scales linearly with the number of
estimated parameters of $\rho(T)$.

In order to properly compare the efficiencies of experimental
gradient-following and global observable tracking methods, it is
necessary to consider the expense associated with reconstruction of
the initial and final density matrices $\rho(0)$ and $\rho(T)$ in the latter case. A variety of
different quantum statistical inference methods have been developed
over the past several years for the estimation of density matrices
on the basis of quantum observations. Like the measurement of the
gradient, these methods are based on multiple observations on
identically prepared copies of the system. If we consider $M$
measurements on identically prepared copies, each measurement is
described by a positive operator-valued measure (POVM). Of interest to us here is the scaling of the
number of measurements required to identify the density matrix
within a given precision, with respect to the Hilbert space
dimension. In most reconstruction techniques, such as quantum
tomography \cite{Ariano1995}, a matrix element of the quantum state
is obtained by averaging its pattern function over data.
In the averaging procedure, the matrix elements are allowed to
fluctuate statistically through negative values, resulting in large
statistical errors. Recently, the method of maximal likelihood estimation (MLE) of quantum states
has received increasing attention due to its greater accuracy \footnote{Other methods for
state reconstruction, such as the maximum entropy method or Bayesian
quantum state identification \cite{Buzek1997}, can also be employed.}. Denoting by $\hat F_i$
the POVM corresponding to the $i$-th observation, the likelihood functional

$$L(\hat \rho) = \prod_{i=1}^N \tr(\hat \rho(0)\hat F_i)$$
describes the probability of obtaining the set of observed outcomes
for a given density matrix $\hat \rho$. This likelihood functional
is maximized over the set of density matrices. An effective
parameterization of $\hat \rho$ is $\hat \rho(0) = \hat T^{\dag} \hat
T$, which guarantees positivity and Hermiticity, and the condition
of unit trace is imposed via a Lagrange multiplier $\lambda$, to
give
$$L(\hat T) = \sum_{i=1}^N \ln \tr (\hat T^{\dag} \hat T \hat F_i) - \lambda \tr(\hat T^{\dag}\hat T).$$
Standard numerical techniques, such as Newton-Raphson or downhill
simplex algorithms, are used to search for the maximum over the
$N^2$ parameters of the matrix $\hat T$. The statistical uncertainty
in density matrix estimates obtained via MLE can quantified by
considering the likelihood to function to represent a probability
distribution on the space of density matrix elements - or, in the
current parameterization, the $N^2$ parameters constituting the
matrix $\hat T$, denoted here by the vector $t$. In the limit of
many measurements, this distribution approaches a Gaussian. The
Fisher information matrix $I = \frac {\partial^2 L}{\partial t
\partial t'}$, which is the variance of the score function that is set to
zero to obtain the MLE estimates, can then be used to quantify the
uncertainties in the parameters. Note that the constraint $\tr(\hat
T^{\dag}\hat T)=1$ implies that the optimization trajectory
maintains orthogonality to $u =\frac {\partial \tr(\hat T^{\dag}\hat
T)}{\partial t}$. Under these conditions it can be shown that the
covariance matrix for $t$ is given by $$V =
I^{-1}-I^{-1}uu^TI^{-1}/u^TI^{-1}u.$$ As such, the associated
uncertainties on the density matrix elements as a function of the
number of measurements can be determined for a given system based on
computer simulations.

Banaszek et al. \cite{Banaszek1999} applied MLE to both discrete and
continuous quantum state estimation, comparing to state tomography.
In the case of continuous variable states, the density matrix was,
of course, truncated. For identical systems, MLE required orders of
magnitude fewer measurements $N$ to reconstruct the state with the
same accuracy. For example, only $50,000$ homodyne data (compared to
$10^8$ for tomography) were required to reconstruct the matrix for a
single-mode radiation field. The number of required measurements was
not highly sensitive to the truncation dimension, since adaptive
techniques can be used to improve efficiency in higher dimensions.
Only $500$ measurements were required to reconstruct the density
matrix of a discrete quantum system, a pair of spin-$1/2$ particles
in the singlet state. Given that accurate estimation of the
gradient requires a similar number of measurements
\cite{Roslund2007}, if MLE is used for state reconstruction, the
additional algorithmic overhead for observable tracking is not
limiting. Quantitative calculations of this overhead will be reported in a
a forthcoming numerical study.

\section{Discussion}

We have presented several global algorithms for the optimization of
quantum observables, based on following globally optimal paths in
the unitary group of dynamical propagators. The most versatile of these algorithms,
orthogonal observable expectation value tracking, aims to simultaneously track a set of
observable expectation value paths consistent with the unitary geodesic path to the target
propagator. The performance of the latter has been compared theoretically to that of
local gradient algorithms. A follow-up paper will report numerical simulations
comparing the efficiencies of the algorithms described herein, across various
families of Hamiltonians.

Although the $\varepsilon(t)$-gradient flow is always the
locally optimal path, its projected path in $\U(N)$ is generally
much longer than those that can be tracked by global algorithms.
The latter often require fewer iterations for
convergence. Of course, in order to
assess the utility of $U$-flow tracking as a practical alternative
to local OCT algorithms, it is necessary to consider the
computational overhead incurred in solving the system of tracking
differential equations at each algorithmic step, which is on the
order of $N^4$-times more costly than solving scalar observable
tracking equations.

For the systems studied, the geodesic track in $\U(N)$ can typically be followed faithfully by matrix tracking algorithms, assuming the system is controllable on the entire unitary group. However, for many Hamiltonians, unitary matrix tracking can
routinely encounter regions of the landscape where the $\textmd{G}$ matrix is
ill-conditioned. In contrast, tracking a vector $\textbf{v}(s)$ of $m$
orthogonal observable expectation values corresponding to this
unitary matrix track encounters such singularities more rarely,
provided $m < N$.

However, the number of observable expectation values tracked is not the only
factor that affects the mean pathlength in $U(N)$. The relationship
between the basis set of operators (spanning the space of measured
observables) and the eigenvalue spectrum of $\rho(0)$ affects the dimension and volume of the subspace of $\U(N)$ that is accessible to the search
trajectory. Indeed, the comparative advantage of employing global observable tracking algorithms versus
local gradient algorithms was found to depend on the spectra of the initial density matrix $\rho(0)$, as predicted based on a geometric analysis
of the map between unitary propagators and associated observable expectation values. In particular,
measurement of the expectation value of a quantum observable provides more information about the
dynamical propagator $U(T)$ when $\rho(0)$ has fewer degenerate eigenvalues, since degeneracies
produce symmetries that result in invariant subspaces over which the unitary propagator can vary without
altering the observable. Thus, for an identical number of measurements, the global path in $\U(N)$ can be tracked with
greater precision for systems with nondegenerate $\rho(0)$.

Clearly, a very important question underlying the efficiency of
global OCT algorithms based on paths in the unitary group is whether
the nonsingularity of $\textmd{G}$ and the assumption of small higher-order
functional derivatives remains valid for more general Hamiltonians
beyond those considered here. In principle, paths in $\U(N)$ that
are longer than the geodesic might be significantly easier to track
if these assumptions were to break down, for particular systems. For
systems where $\textmd{G}$ is almost always close to singular, global
tracking algorithms on $\U(N)$ may not be viable, even in the
presence of error correction. In these cases one would expect global
observable tracking to be preferable to matrix tracking algorithms,
since the system can "choose" which of the infinite number of
degenerate paths in $\U(N)$ it follows.

This study, and the associated forthcoming numerical simulations, sets the stage for experimental testing of its prediction that global observable
tracking algorithms will display advantages compared to the gradient. As discussed above, an important question for future study is how to implement global observable control algorithms
experimentally, and whether nonideal conditions (noise, decoherence) in the laboratory will obscure some of its
predicted advantages. In particular, the \textit{effective} degeneracies of $\rho(0)$ and $\Theta$ - e.g., whether the populations of the various
levels in a mixed state are sufficiently high to permit accurate determination of the associated unitary
propagators in the presence of noise -  become very important.

Besides the perceived advantage of global observable control algorithms, they may offer
a means of assessing the search effort and search complexity inherent in quantum control
problems in a universal manner, since they are more system independent than local
gradient-based algorithms. As shown in a separate numerical study comparing these algorithms, over several families of related Hamiltonians, the variance of the convergence time of the $\varepsilon(t)$-gradient flow is significantly
greater than that of global observable or unitary matrix tracking.
Given that its convergence is also faster, unitary tracking may offer an approach to setting upper bounds on the scaling of the
time required for quantum observable control optimizations, as a
function of system size.

\bibliography{INTEGRABLE}

\begin{thebibliography}{27}
\expandafter\ifx\csname natexlab\endcsname\relax\def\natexlab#1{#1}\fi
\expandafter\ifx\csname bibnamefont\endcsname\relax
  \def\bibnamefont#1{#1}\fi
\expandafter\ifx\csname bibfnamefont\endcsname\relax
  \def\bibfnamefont#1{#1}\fi
\expandafter\ifx\csname citenamefont\endcsname\relax
  \def\citenamefont#1{#1}\fi
\expandafter\ifx\csname url\endcsname\relax
  \def\url#1{\texttt{#1}}\fi
\expandafter\ifx\csname urlprefix\endcsname\relax\def\urlprefix{URL }\fi
\providecommand{\bibinfo}[2]{#2}
\providecommand{\eprint}[2][]{\url{#2}}

\bibitem[{\citenamefont{Chakrabarti and Rabitz}(2007)}]{Raj2007}
\bibinfo{author}{\bibfnamefont{R.}~\bibnamefont{Chakrabarti}} \bibnamefont{and}
  \bibinfo{author}{\bibfnamefont{H.}~\bibnamefont{Rabitz}},
  \bibinfo{journal}{International Reviews in Physical Chemistry}
  \textbf{\bibinfo{volume}{26}} (\bibinfo{year}{2007}).

\bibitem[{\citenamefont{Rabitz et~al.}(2005)\citenamefont{Rabitz, Hsieh, and
  Rosenthal}}]{RabMik2005}
\bibinfo{author}{\bibfnamefont{H.}~\bibnamefont{Rabitz}},
  \bibinfo{author}{\bibfnamefont{M.}~\bibnamefont{Hsieh}}, \bibnamefont{and}
  \bibinfo{author}{\bibfnamefont{C.}~\bibnamefont{Rosenthal}},
  \bibinfo{journal}{Physical Review A} \textbf{\bibinfo{volume}{72}},
  \bibinfo{pages}{52337} (\bibinfo{year}{2005}).

\bibitem[{\citenamefont{Hsieh et~al.}(2006)\citenamefont{Hsieh, Wu, and
  Rabitz}}]{Mike2006a}
\bibinfo{author}{\bibfnamefont{M.}~\bibnamefont{Hsieh}},
  \bibinfo{author}{\bibfnamefont{R.}~\bibnamefont{Wu}}, \bibnamefont{and}
  \bibinfo{author}{\bibfnamefont{H.}~\bibnamefont{Rabitz}},
  \bibinfo{journal}{J. Chem. Phys.} \textbf{\bibinfo{volume}{51}},
  \bibinfo{pages}{204107} (\bibinfo{year}{2006}).

\bibitem[{\citenamefont{Roslund and Rabitz}(2007)}]{Roslund2007}
\bibinfo{author}{\bibfnamefont{J.}~\bibnamefont{Roslund}} \bibnamefont{and}
  \bibinfo{author}{\bibfnamefont{H.}~\bibnamefont{Rabitz}},
  \bibinfo{journal}{To be submitted}  (\bibinfo{year}{2007}).

\bibitem[{\citenamefont{Khaneja and Glaser}(2001)}]{Khaneja2001}
\bibinfo{author}{\bibfnamefont{N.}~\bibnamefont{Khaneja}} \bibnamefont{and}
  \bibinfo{author}{\bibfnamefont{S.}~\bibnamefont{Glaser}},
  \bibinfo{journal}{Chem. Phys.} \textbf{\bibinfo{volume}{267}},
  \bibinfo{pages}{11} (\bibinfo{year}{2001}).

\bibitem[{\citenamefont{Khaneja et~al.}(2001)\citenamefont{Khaneja, Brockett,
  and Glaser}}]{Khaneja2002a}
\bibinfo{author}{\bibfnamefont{N.}~\bibnamefont{Khaneja}},
  \bibinfo{author}{\bibfnamefont{R.}~\bibnamefont{Brockett}}, \bibnamefont{and}
  \bibinfo{author}{\bibfnamefont{S.}~\bibnamefont{Glaser}},
  \bibinfo{journal}{Phys. Rev. A} \textbf{\bibinfo{volume}{63}},
  \bibinfo{pages}{032308} (\bibinfo{year}{2001}).

\bibitem[{\citenamefont{Wu and Rabitz}(2007)}]{Wu2007}
\bibinfo{author}{\bibfnamefont{R.}~\bibnamefont{Wu}} \bibnamefont{and}
  \bibinfo{author}{\bibfnamefont{H.}~\bibnamefont{Rabitz}},
  \bibinfo{journal}{in preparation}  (\bibinfo{year}{2007}).

\bibitem[{\citenamefont{Jezek et~al.}(2003)\citenamefont{Jezek, Fiurasek, and
  Hradil}}]{Hradil2003}
\bibinfo{author}{\bibfnamefont{M.}~\bibnamefont{Jezek}},
  \bibinfo{author}{\bibfnamefont{J.}~\bibnamefont{Fiurasek}}, \bibnamefont{and}
  \bibinfo{author}{\bibfnamefont{Z.}~\bibnamefont{Hradil}},
  \bibinfo{journal}{Phys. Rev. A} \textbf{\bibinfo{volume}{68}},
  \bibinfo{pages}{012305} (\bibinfo{year}{2003}).

\bibitem[{\citenamefont{Mohseni et~al.}(2007)\citenamefont{Mohseni, Rezakhani,
  and Lidar}}]{Lidar2007}
\bibinfo{author}{\bibfnamefont{M.}~\bibnamefont{Mohseni}},
  \bibinfo{author}{\bibfnamefont{A.~T.} \bibnamefont{Rezakhani}},
  \bibnamefont{and} \bibinfo{author}{\bibfnamefont{D.~A.} \bibnamefont{Lidar}},
  \bibinfo{journal}{arXiv:quant-ph/0702131v1}  (\bibinfo{year}{2007}).

\bibitem[{\citenamefont{Malley and Hornstein}(1993)}]{Malley1993}
\bibinfo{author}{\bibfnamefont{J.}~\bibnamefont{Malley}} \bibnamefont{and}
  \bibinfo{author}{\bibfnamefont{J.}~\bibnamefont{Hornstein}},
  \bibinfo{journal}{Stat. Sci.} \textbf{\bibinfo{volume}{8}},
  \bibinfo{pages}{433} (\bibinfo{year}{1993}).

\bibitem[{\citenamefont{Rothman et~al.}(2005)\citenamefont{Rothman, Ho, and
  Rabitz}}]{Rothman2005}
\bibinfo{author}{\bibfnamefont{A.}~\bibnamefont{Rothman}},
  \bibinfo{author}{\bibfnamefont{T.}~\bibnamefont{Ho}}, \bibnamefont{and}
  \bibinfo{author}{\bibfnamefont{H.}~\bibnamefont{Rabitz}},
  \bibinfo{journal}{Phys. Rev. A} \textbf{\bibinfo{volume}{72}},
  \bibinfo{pages}{023416} (\bibinfo{year}{2005}).

\bibitem[{\citenamefont{Rothman
  et~al.}(2006{\natexlab{a}})\citenamefont{Rothman, Ho, and
  Rabitz}}]{Rothman2006a}
\bibinfo{author}{\bibfnamefont{A.}~\bibnamefont{Rothman}},
  \bibinfo{author}{\bibfnamefont{T.}~\bibnamefont{Ho}}, \bibnamefont{and}
  \bibinfo{author}{\bibfnamefont{H.}~\bibnamefont{Rabitz}},
  \bibinfo{journal}{J. Chem. Phys.} \textbf{\bibinfo{volume}{123}},
  \bibinfo{pages}{134104} (\bibinfo{year}{2006}{\natexlab{a}}).

\bibitem[{\citenamefont{Rothman
  et~al.}(2006{\natexlab{b}})\citenamefont{Rothman, Ho, and
  Rabitz}}]{Rothman2006b}
\bibinfo{author}{\bibfnamefont{A.}~\bibnamefont{Rothman}},
  \bibinfo{author}{\bibfnamefont{T.}~\bibnamefont{Ho}}, \bibnamefont{and}
  \bibinfo{author}{\bibfnamefont{H.}~\bibnamefont{Rabitz}},
  \bibinfo{journal}{Phys. Rev. A} \textbf{\bibinfo{volume}{73}},
  \bibinfo{pages}{053401} (\bibinfo{year}{2006}{\natexlab{b}}).

\bibitem[{\citenamefont{Ho and Rabitz}(2006)}]{HoRab2007a}
\bibinfo{author}{\bibfnamefont{T.}~\bibnamefont{Ho}} \bibnamefont{and}
  \bibinfo{author}{\bibfnamefont{H.}~\bibnamefont{Rabitz}},
  \bibinfo{journal}{J. Photochem. Photobiol. A} \textbf{\bibinfo{volume}{180}},
  \bibinfo{pages}{226} (\bibinfo{year}{2006}).

\bibitem[{\citenamefont{Dominy and Rabitz}(2007)}]{Dominy2007}
\bibinfo{author}{\bibfnamefont{J.}~\bibnamefont{Dominy}} \bibnamefont{and}
  \bibinfo{author}{\bibfnamefont{H.}~\bibnamefont{Rabitz}},
  \bibinfo{journal}{In preparation}  (\bibinfo{year}{2007}).

\bibitem[{\citenamefont{Brockett}(1991)}]{Brockett1991}
\bibinfo{author}{\bibfnamefont{R.}~\bibnamefont{Brockett}},
  \bibinfo{journal}{Linear Alg Appl.} \textbf{\bibinfo{volume}{146}},
  \bibinfo{pages}{79} (\bibinfo{year}{1991}).

\bibitem[{\citenamefont{Ho}(2007)}]{HoRab2007b}
\bibinfo{author}{\bibfnamefont{T.}~\bibnamefont{Ho}}, \bibinfo{journal}{In
  preparation}  (\bibinfo{year}{2007}).

\bibitem[{\citenamefont{Bloch et~al.}(1992)\citenamefont{Bloch, Brockett, and
  Ratiu}}]{Bloch1992}
\bibinfo{author}{\bibfnamefont{A.}~\bibnamefont{Bloch}},
  \bibinfo{author}{\bibfnamefont{R.}~\bibnamefont{Brockett}}, \bibnamefont{and}
  \bibinfo{author}{\bibfnamefont{T.}~\bibnamefont{Ratiu}},
  \bibinfo{journal}{Commun. Math. Phys.} \textbf{\bibinfo{volume}{147}},
  \bibinfo{pages}{57} (\bibinfo{year}{1992}).

\bibitem[{\citenamefont{Helmke and Moore}(1994)}]{Helmke1994}
\bibinfo{author}{\bibfnamefont{U.}~\bibnamefont{Helmke}} \bibnamefont{and}
  \bibinfo{author}{\bibfnamefont{J.}~\bibnamefont{Moore}},
  \emph{\bibinfo{title}{Optimization and dynamical systems}}
  (\bibinfo{publisher}{Springer-Verlag}, \bibinfo{address}{London},
  \bibinfo{year}{1994}).

\bibitem[{\citenamefont{Faybusovich}(1991)}]{Faybu1991}
\bibinfo{author}{\bibfnamefont{L.}~\bibnamefont{Faybusovich}},
  \bibinfo{journal}{Physica D} \textbf{\bibinfo{volume}{23}},
  \bibinfo{pages}{309} (\bibinfo{year}{1991}).

\bibitem[{\citenamefont{Bloch}(1990)}]{Bloch1990}
\bibinfo{author}{\bibfnamefont{A.}~\bibnamefont{Bloch}},
  \bibinfo{journal}{Contemp. Math.} \textbf{\bibinfo{volume}{114}},
  \bibinfo{pages}{77} (\bibinfo{year}{1990}).

\bibitem[{\citenamefont{Bloch}(1995)}]{Bloch1995}
\bibinfo{author}{\bibfnamefont{A.}~\bibnamefont{Bloch}},
  \emph{\bibinfo{title}{Hamiltonian and Gradient Flows, Algorithms and
  Control}}, vol.~\bibinfo{volume}{3} of \emph{\bibinfo{series}{Fields
  Institute Commmunications}} (\bibinfo{publisher}{Oxford University Press},
  \bibinfo{address}{Oxford}, \bibinfo{year}{1995}).

\bibitem[{\citenamefont{Babelon et~al.}(2003)\citenamefont{Babelon, Bernard,
  and Talon}}]{Babelon2003}
\bibinfo{author}{\bibfnamefont{O.}~\bibnamefont{Babelon}},
  \bibinfo{author}{\bibfnamefont{D.}~\bibnamefont{Bernard}}, \bibnamefont{and}
  \bibinfo{author}{\bibfnamefont{M.}~\bibnamefont{Talon}},
  \emph{\bibinfo{title}{Introduction to classical integrable systems}},
  vol.~\bibinfo{volume}{60} of \emph{\bibinfo{series}{Cambridge Monographs on
  Mathematical Physics}} (\bibinfo{publisher}{Cambridge},
  \bibinfo{address}{Cambridge}, \bibinfo{year}{2003}).

\bibitem[{\citenamefont{Chakrabarti et~al.}(2007)\citenamefont{Chakrabarti, Wu,
  and Rabitz}}]{RajWu2007}
\bibinfo{author}{\bibfnamefont{R.}~\bibnamefont{Chakrabarti}},
  \bibinfo{author}{\bibfnamefont{R.}~\bibnamefont{Wu}}, \bibnamefont{and}
  \bibinfo{author}{\bibfnamefont{H.}~\bibnamefont{Rabitz}},
  \bibinfo{journal}{In preparation}  (\bibinfo{year}{2007}).

\bibitem[{\citenamefont{D'Ariano}(1995)}]{Ariano1995}
\bibinfo{author}{\bibfnamefont{G.}~\bibnamefont{D'Ariano}},
  \bibinfo{journal}{Phys. Rev. A} \textbf{\bibinfo{volume}{52}},
  \bibinfo{pages}{R1801} (\bibinfo{year}{1995}).

\bibitem[{\citenamefont{Buzek et~al.}(1997)\citenamefont{Buzek, Derka, Adam,
  and Knight}}]{Buzek1997}
\bibinfo{author}{\bibfnamefont{V.}~\bibnamefont{Buzek}},
  \bibinfo{author}{\bibfnamefont{R.}~\bibnamefont{Derka}},
  \bibinfo{author}{\bibfnamefont{G.}~\bibnamefont{Adam}}, \bibnamefont{and}
  \bibinfo{author}{\bibfnamefont{P.}~\bibnamefont{Knight}},
  \bibinfo{journal}{Ann. of Phys.} \textbf{\bibinfo{volume}{266}},
  \bibinfo{pages}{454} (\bibinfo{year}{1997}).

\bibitem[{\citenamefont{Banaszek et~al.}(1999)\citenamefont{Banaszek, D'Ariano,
  Paris, and Sacchi}}]{Banaszek1999}
\bibinfo{author}{\bibfnamefont{K.}~\bibnamefont{Banaszek}},
  \bibinfo{author}{\bibfnamefont{G.}~\bibnamefont{D'Ariano}},
  \bibinfo{author}{\bibfnamefont{M.}~\bibnamefont{Paris}}, \bibnamefont{and}
  \bibinfo{author}{\bibfnamefont{M.}~\bibnamefont{Sacchi}},
  \bibinfo{journal}{Physical Review A} \textbf{\bibinfo{volume}{61}},
  \bibinfo{pages}{010304} (\bibinfo{year}{1999}).

\end{thebibliography}

\end{document}